\documentclass[twocolumn]{aastex63}

\shorttitle{The Kinematic Substructure of Local Dark Matter}
\shortauthors{Zhu et al.}
\usepackage{fancyhdr}
\usepackage{amsmath}
\usepackage{hyperref}
\hypersetup{
    colorlinks=true,
    linkcolor=blue,
    filecolor=blue,      
    urlcolor=blue,
    citecolor=blue,
}
\usepackage{xcolor}
\allowdisplaybreaks
\usepackage{graphicx}
\usepackage{appendix}
\usepackage{lineno}
\makeatletter
\newcommand\footnoteref[1]{\protected@xdef\@thefnmark{\ref{#1}}\@footnotemark}
\makeatother
\usepackage[section]{placeins}
\usepackage{amssymb}
\usepackage{multirow}
\usepackage{booktabs}
\usepackage{float} 
\usepackage{subfigure}
\usepackage{natbib}
\usepackage[marginal]{footmisc}
\renewcommand{\thefootnote}{\arabic{footnote}}
\renewcommand{\thempfootnote}{\arabic{mpfootnote}}
\newcommand{\gaia}{{\it Gaia}}
\newcommand{\kms}{km\,s$^{-1}$}
\newcommand{\teff}{$T_{\rm{eff}}$/K}
\newcommand{\logg}{log\,$g$}
\newcommand{\feh}{\rm{[Fe/H]}}

\newcommand{\kpc}{kpc}
\newcommand{\vlos}{$v_{\rm{los}}$}
\graphicspath{{./}{figures/}}
\begin{document}
\title{Deciphering the Kinematic Substructure of Local Dark Matter with LAMOST K Giants}
\email{jtshen@sjtu.edu.cn}
\author{Hai Zhu\footnotemark[8]}
\footnotetext[8]{Contributed equally to this work.}
\affiliation{School of Physics and Astronomy, Shanghai Jiao Tong University, Key Laboratory for Particle Astrophysics
and Cosmology (MoE), Shanghai Key Laboratory for Particle Physics and Cosmology, Shanghai 200240, People's Republic of China}
\author{Rui Guo\footnotemark[8]}
\affiliation{School of Physics and Astronomy, Shanghai Jiao Tong University, Key Laboratory for Particle Astrophysics
and Cosmology (MoE), Shanghai Key Laboratory for Particle Physics and Cosmology, Shanghai 200240, People's Republic of China}
\author{Juntai Shen}
\affiliation{School of Physics and Astronomy, Shanghai Jiao Tong University, Key Laboratory for Particle Astrophysics
and Cosmology (MoE), Shanghai Key Laboratory for Particle Physics and Cosmology, Shanghai 200240, People's Republic of China}
\author{Jianglai Liu}
\affiliation{New Cornerstone Science Laboratory, Tsung-Dao Lee Institute, Shanghai Jiao Tong University, Shanghai 200240, People's Republic of China}
\affiliation{School of Physics and Astronomy, Shanghai Jiao Tong University, Key Laboratory for Particle Astrophysics
and Cosmology (MoE), Shanghai Key Laboratory for Particle Physics and Cosmology, Shanghai 200240, People's Republic of China}
\affiliation{Shanghai Jiao Tong University Sichuan Research Institute, Chengdu 610213, People's Republic of China}
\author{Chao Liu}
\affiliation{Key Laboratory of Space Astronomy and Technology, National Astronomical Observatories, Chinese Academy of Sciences, Beijing 100101, People's Republic of China}
\affiliation{Institute for Frontiers in Astronomy and Astrophysics, Beijing Normal University, Beijing 102206, People's Republic of China}
\author{Xiang-Xiang Xue}
\affiliation{CAS Key Laboratory of Optical Astronomy, National Astronomical Observatories, Chinese Academy of Sciences, Beijing 100101, People's Republic of China}
\affiliation{Institute for Frontiers in Astronomy and Astrophysics, Beijing Normal University, Beijing 102206, People's Republic of China}
\author{Lan Zhang}
\affiliation{CAS Key Laboratory of Optical Astronomy, National Astronomical Observatories, Chinese Academy of Sciences, Beijing 100101, People's Republic of China}
\author{Shude Mao}
\affiliation{Department of Astronomy and Tsinghua Centre for Astrophysics, Tsinghua University, Beijing 100084, People's Republic of China}
\begin{abstract}
Numerical simulations indicate that correlations exist between the velocity distributions of stars and dark matter (DM). We study the local DM velocity distribution based on these correlations. We select K giants from LAMOST DR8 cross-matched with {\gaia} DR3, which have robust measurements of velocity and metallicity, and separate them into the disk, halo substructure and isotropic halo components in the chemodynamical space utilizing the Gaussian Mixture Model. The substructure component is highly radially anisotropic, and possibly related to the \gaia-Enceladus-Sausage (GES) merger event, while the isotropic halo component is accreted from the earliest mergers following the Maxwell-Boltzmann distribution (Standard Halo Model, SHM). We find that the GES-like substructure contributes $\sim85\%$ of the local nondisk stars in the Solar neighbourhood, which is nearly invariant when applying different volume cuts or additional angular momentum constraints. Utilizing the metallicity-stellar mass relation and the stellar mass-halo mass relation, we find that $\sim25_{-15}^{+24}\%$ of local DM is in the kinematic substructure. Combined with the stellar distributions of nondisk components, we modify the heliocentric velocity distribution of local DM. It shifts to a lower speed with a sharper peak compared to the SHM, and updates the detection limits of DM direct detection experiments. We discuss extensively the degeneracies present in the GMM fitting and propose that more kinematic and chemical information such as $\alpha$ abundance could help to break the degeneracy in the future.  Our work confirms that the local DM velocity distribution deviates significantly from the SHM, and needs to be properly accounted for in the DM detection experiments.
 
\end{abstract}


\section{Introduction}
\label{sec:intro}

In the $\Lambda$CDM paradigm, the Milky Way halo is built up from hierarchical merger of smaller satellites \citep{WR_1978}. Depending on the merger time and properties of the satellites, such as the mass ratio and the satellite orbits, the satellite remnants that construct the Milky Way halo at present day can be separated into different types. The earliest merger has the longest time to virialize after the dynamical relaxation, which results in a smooth and isotropic distribution. This isotropic component can be described by the Standard Halo Model (SHM) \citep{Ostriker_1974,Bahcall_1980,Caldwell_1981} that follows the Maxwell-Boltzmann velocity distribution \citep{Drukier_1986,Freese_1988}. The more recent merger events usually do not have enough time to fully mix into the main halo even in the configuration space. In particular, the outer part of the satellite is often stripped away to form a tidal stream. The stellar streams have been widely used to study the Milky Way mass distribution and its dark matter (DM) halo shape \citep[e.g.][]{Malhan_2019,Vasiliev_2020,Hattori_2021,Ibata_2021,Koposov_2023}. An intermediate merger between ancient and recent ones leaves substructures called the ``debris flows'', which are well mixed with the main halo in the configuration space, but showing structures in the kinematic space \citep{Helmi_1999,Gomez_2010}.

The most important merger event that built up the Milky Way might be the \gaia-Enceladus-Sausage (GES) \citep{Helmi_2018,Belokurov_2018b}, which remains a sausage-like structure in the radial and azimuthal $v_{r}$-$v_{\phi}$ velocity space. Similar sausage-like halo stars are also found in an earlier work in the observation \citep{Chiba_2000} and can be explained by a polar-orbit satellite accretion \citep{Brook_2003}. The progenitor of the GES is thought to be a large dwarf galaxy with a highly eccentric orbit and a stellar mass on the order of $10^{9} - 10^{10}$ M$_{\odot}$ \citep{Vincenzo_2019,Deason_2019,Fattahi_2019,Koppelman_2020a} and merged with the Milky Way around 8$-$10 Gyr ago \citep{Sahlholdt_2019,Bignone_2019,Bonaca_2020}. The chemodynamical studies indicate that the GES stars have a metallicity peak of {\feh} around $-1.4$ to $-1.2$ dex \citep{Sahlholdt_2019,Gallart_2019,Das_2020,Feuillet_2020,Naidu_2020}. Due to its large mass, GES strongly influenced the radial kinematic profiles of the stellar halo. This highly radially anisotropic component exists in the local stellar halo (within $\sim$ 10~{\kpc} from the Sun) and extends out to a Galactocentric radius $r_{gc} \sim $ 25~{\kpc} \citep{Lancaster_2019,Bird_2019,Bird_2021,Wu_2022}. \citet{Deason_2018} showed that the apocenter of the GES debris stars is about $20-25~${\kpc}, which might be related to the break radius of the density profile of the Galactic stellar halo \citep{Watkins_2009,Sesar_2011,Pila_Diez_2015,Xue_2015}. \citet{Wu_2022} found that the GES contributes about $41\%-74\%$ of the inner ($r_{gc} < 30$~{\kpc}) stellar halo. In the Solar neighbourhood, \citet{Necib_2019} showed that the GES stars contribute $60\%-80\%$ of the nondisk components of main sequence stars in the SDSS DR9 sample, with a peak metallicity of $\feh = -1.4$ dex. 

In the DM direct detection experiments, both the density and velocity distributions of the local DM are important inputs to calculate the differential scattering rates and the detection limit curves \citep[e.g.][]{Lewin_1996,Frandsen_2012,Bhattacharjee_2013,Fairbairn_2013,Green_2017}. The local DM density has been widely measured utilizing the local stellar kinematics and the Galactic rotation curve \citep[e.g.][]{Salucci_2010,Fabrizio_2013,Xia_2016,Hagen_2018,Huang_2016,de_Salas_2019,Eilers_2019,Guo_2020,Guo_2022,Zhou_2023,Ou_2024}, while the DM velocity distribution is usually assumed to follow the Maxwell-Boltzmann distribution \citep{Drukier_1986, Freese_1988}. However, the prior merger events and the recent interactions with satellites could change the local DM velocity distribution \citep{Necib_2019,Donlon_2022,Besla_2019,Maity_2023,smithorlik_2023,Donaldson_2022,Evans_2019}. \citet{Evans_2019} found that the strongly radially anisotropic GES may increase the fraction of high velocity tail relative to the SHM, leading to a harder kinetic energy distribution of incoming DM particles in detection experiments and hence a strong exclusion limit for lower mass DM. Nevertheless, \citet{Necib_2019} utilized the stellar velocity distribution to trace the DM velocity distribution, and found a contrary influence of the GES on the DM detection experiments. The differences mainly result from the contribution of the GES to the local DM and the profiles of its velocity distribution. This tension requests for a more detailed study on the exact stellar velocity distribution of the halo and substructures.

Some simulations of Milky-Way-like galaxies show that there are correlations between the velocity distribution of stars and the accreted DM particles \citep{Herzog_2018a,Lisanti_2015,Necib_2019b}, i.e. the stars and DM accreted from the same merger satellite may share similar velocity distribution\footnote{The stars formed in the accreted gas are not included, as their kinematic information could be different after the star formation.}. This allows one to trace the DM velocity distribution with halo stars. The velocity distribution of the DM accreted from the oldest mergers forming the virialized halo can be well traced by metal-poor stars in the \texttt{Eris} hydrodynamic simulations \citep{Herzog_2018a}, which motivates the first study using the RAVE-TGAS data to recover the velocity distribution of the local relaxed DM component in \citet{Herzog_2018b}. The DM accreted from younger mergers, which remained as debris flows and streams, are associated with more metal-rich stars, as illustrated in the \texttt{Via Lactea} simulations and \texttt{FIRE-2} simulations \citep{Lisanti_2015,Necib_2019b}. The correlation between DM and star particles found in the \texttt{FIRE-2} simulations holds for two simulated hosts with different merger histories, motivating the study of the kinematic substructure of DM in the Solar neighbourhood in \cite{Necib_2019}. In this framework the stellar metallicity distribution is crucial and requires a large sample of stars with accurate metallicity measurement. 
\cite{Necib_2019} found their results are invariant for samples with different vertical cuts down to $|z|>2.5~${\kpc}, which supports the extrapolation of the results to the Galactic plane. Even lower regions are not investigated because of the large contribution of disk stars. Besides, though their corner plots show some degeneracy for each component, the degeneracy between the different components and their relative fractions is not fully explored. They also pointed out that the fractional contributions pertain only to their data set and that metallicity biases can potentially affect the extrapolation to the Solar vicinity.

In this work we attempt to provide an independent check on the contribution of the GES to the stellar halo and explore the region closer to the Galactic plane utilizing another stellar tracer, K giants selected from LAMOST \citep{Cui_2012,Zhao_2012}. These K giants have high-quality measurements of line-of-sight velocity and metallicity, and are cross-matched with {\gaia} DR3 \citep{GaiaDR3} for proper motions measurements. More importantly, our K giants sample has accurate distance estimation. We follow the study of \citet{Necib_2019} to extract information of different components in the chemodynamical space applying the Gaussian Mixture Model (GMM) method. The potential model degeneracies are also analysed in depth in our work. Assuming that there exists correlations between the DM and halo stars, we obtain the velocity distribution of the local DM considering the kinematic substructure of the GES, and then to investigate its influence on the DM direct detection experiments.


This paper is organized as follows: In Section \ref{Sec2: Data Analysis}, we discuss the selection of the K giants sample and the likelihood calculation procedure of the GMM method. In Section \ref{Sec3: Stellar Distribution in the Chemodynamical Space}, we present the distributions of different stellar components from our best-fit model and explore the variation due to different spatial volume cuts of the samples. To reduce the dominance of the disk stars, we apply an angular momentum cut to the samples, and extend the GMM fitting to halo stars near the disk in Section \ref{Sec4: Angular Momentum Constrain}. The application of the results to the DM velocity distribution and direct detection experiments are shown in Section \ref{Sec5: Dark Matter Distribution and Detection Limits}. Discussions and summary are presented in Section \ref{Sec6: Discussion} and Section \ref{Sec7: Conclusion}, respectively.

\section{Data Analysis}
\label{Sec2: Data Analysis}
\subsection{Sample Selection}
\label{ssec: samp_selection}

\begin{figure}
\centering
\includegraphics[width=\columnwidth]{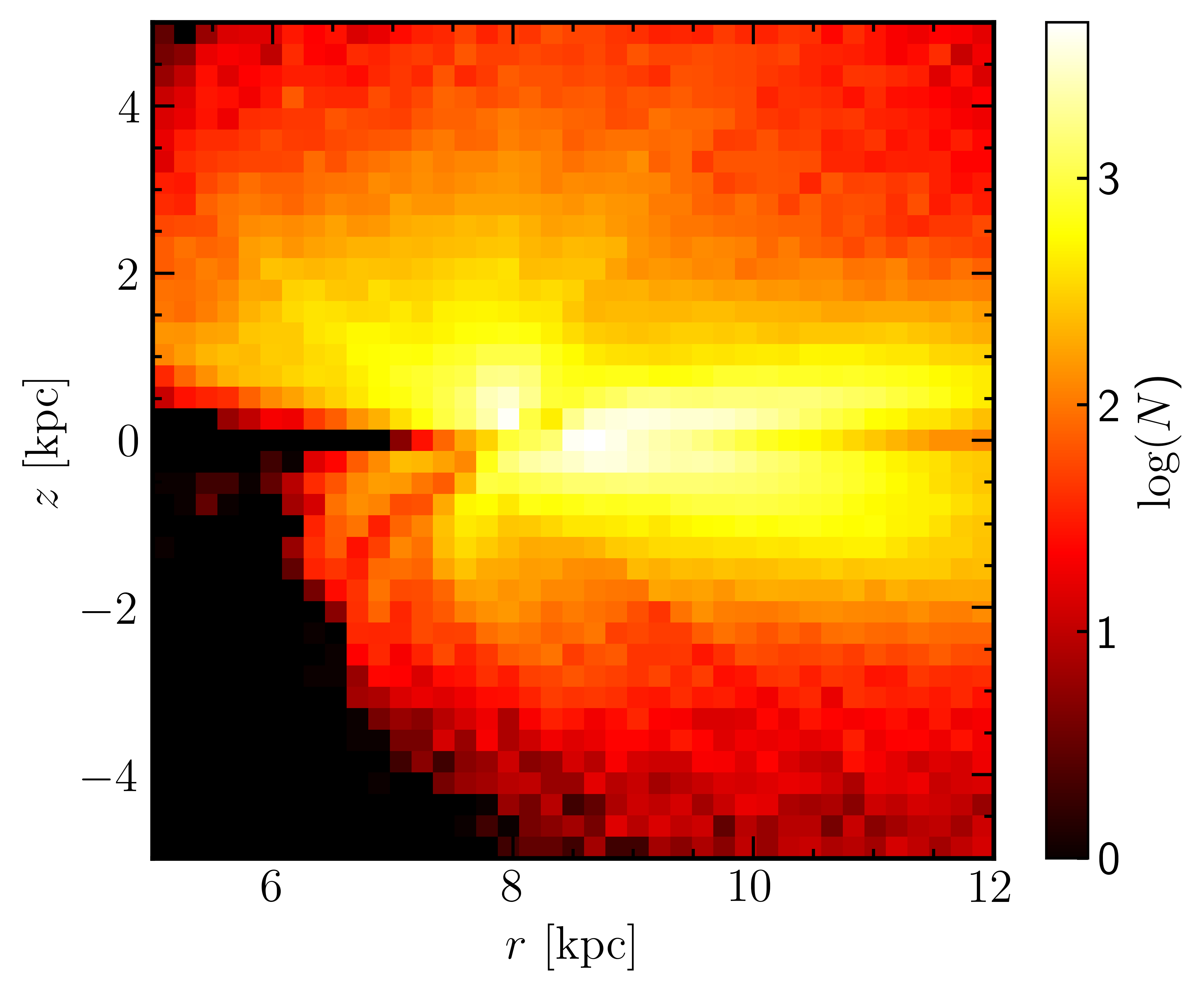}
\caption{Spatial distribution of the LAMOST DR8 $\times$ {\gaia} DR3 sample in terms of Galactocentric radius $r$ and vertical distance from the Galactic disk $z$. Stars within $r\in[7.5, 10.0]$~kpc and $|z|>$ 2.5~kpc are utilized in the following chemodynamical analyses.}
\label{r_z}
\end{figure}

In order to study the kinematic substructure in the local stellar halo, we consider the K giants selected from LAMOST Data Release 8 (DR8). The selection criteria of the LAMOST K giants are based on the surface gravity \logg\ and the effective temperature $T_{\rm{eff}}$ according to \citet{Liu_2014}, i.e. \logg\ $\leq4$ when $4600 \leq$\teff$\leq5600$ or \logg\ $\leq3.5$ when 4000 $\leq$\teff$\leq4600$. We cross-match the K giant sample with {\gaia} DR3 with a radius $1^{\prime\prime}$ in order to obtain accurate kinematic information. LAMOST provides accurate metallicity ({\feh}) and line-of-sight velocity ({\vlos}) measurements from the spectra, while {\gaia} provides accurate proper motions. The cross-matched sample contains 612,136 K giants. The LAMOST {\vlos} has a systematic offset of 5.34 {\kms} for K giants relative to that from {\gaia}, which was investigated \citep{Tian_2015,Ding_2021} and has been corrected in this work. Due to the large uncertainty in distance derived from parallax for distant stars, the distance ($d$) is estimated with the Bayesian method described in \citet{Xue_2014}. We further exclude stars with distance error $\delta d/d>0.3$ and absolute spherical velocity $|\boldsymbol{v}| >400$ {\kms}. This results in a sample containing 438,466 K giants, 18,484 of which are potential halo stars within the Solar neighbourhood ($r\in[7.5,10]$~{\kpc} \& $|z|>2.5$~{\kpc}). This K giant sample has a mean uncertaintiy of 20\% in the distance, 4.4~{\kms} in {\vlos} and 0.06~dex in the metallicity. We provide a detailed comparison of some characteristic uncertainties between the samples utilized in previous works and ours in Table~\ref{table2} in the Appendix. The metallicity has a floor at $-2.5$~dex due to the limit of LAMOST spectroscopic pipeline \citep{Wu_2011,Luo_2015} which has few template stars with $\feh < -2.5$, and the limitation of spectroscopic data that the very weak variation in LAMOST low resolution spectra near $\feh =-2.5$ makes it difficult to detect stars with lower metallicity. Nevertheless, the fraction of stars below this metallicity floor of $-2.5$ is estimated to be small ($< 1\%$) \citep{Schorck_2009,Li_2010}, leading to a negligible impact to our metallicity distribution and results.

The coordinates system adopted in this paper is the Galactocentric spherical coordinates ($r, \theta, \phi$), pointing outwards, to the north Galactic pole and to the Galactic rotation, respectively. We adopt a Solar distance to the Galactic center of $R_{0}=8.122$~{\kpc} \citep{2018Gravity}, and a Solar vertical height of $z_{\odot}=20.8$ pc \citep{Bennett_2018}, with a Solar motion of ($+12.9,+245.6,+7.78$)~{\kms} in the Galactocentric Cartesian coordinates \citep{Reid_2004,2018Gravity,Drimmel_2018}. We utilize the {\tt python} package \textbf{astropy} \citep{Price-Whelan_2018} to transform the astrometric measurements of proper motions ($\mu_{\alpha},\mu_{\delta}$) and {\vlos} to the kinematic information in the spherical coordinates. The measurement errors are propagated to the uncertainties in the velocities using a
Monte Carlo sampling with the {\tt python} package {\tt pyia} \citep{PW_2018}. Fig.~\ref{r_z} shows the spatial distribution of the sample we used. We focus mainly on the region of $r\in[7.5, 10.0]$~kpc and $|z|>$2.5 kpc, and explore the regions closer to the Galactic plane in Section \ref{Sec4: Angular Momentum Constrain}.

\subsection{Procedure of Likelihood Calculation}
\label{Procedure of Likelihood Calculation}

The stellar sample can be separated into three components, i.e. the halo, the substructure and the disk, according to the stellar chemodynamical information \citep{Necib_2019}. The halo component represents the isotropic and metal-poor halo contributed by the oldest mergers, while the substructure is a radially anisotropic halo component mainly contributed by the GES merger event. The $v_{r}$ distribution of the substructure shows a symmetric bimodal Gaussian distribution, due to the highly radial merger trajectory. The disk component contains more metal-rich stars born in-situ with strong azimuthal rotation. As these components show different chemical and kinematic properties, we utilize the GMM to identify different components with a Bayesian estimation \citep{Lancaster_2019,Necib_2019}, in the ($v_{r},\ v_{\theta},\ v_{\phi},\ \rm\feh$) chemodynamical space.

\begin{table}[]
\begin{center}
\caption{Priors of 35 free parameters.}
\label{table1}
\setlength{\tabcolsep}{1.1mm}{
\begin{tabular}{ccccc}
\toprule
\toprule
\multicolumn{1}{c}{\multirow{2}{*}{Parameter}}   & \multicolumn{1}{c}{\multirow{2}{*}{Type}} &\multicolumn{3}{c}{Priors} \\
\cline{3-5}
    & &Disk & Halo& Substructure\\
\hline
$\mu_{r}$ &   Linear  & $[-70,\ 70]$   &$[-70,\ 70]$&$[0,\ 250]$\\
$\mu_{\theta}$ &   Linear  & $[-70,\ 70]$   &$[-70,\ 70]$&$[-70,\ 70]$\\
$\mu_{\phi}$ &   Linear  & $[0,\ 300]$   &$[-70,\ 70]$&$[-70,\ 70]$\\
$\sigma_{r,\theta,\phi}$ &   Linear  & $[0,\ 200]$   &$[0,\ 200]$&$[0,\ 200]$\\
$\rho_{r\theta,r\phi,\theta\phi}$ &   Linear  & $[-1,\ 1]$   &$[-1,\ 1]$&$[-1,\ 1]$\\
$\mu_{\mathrm{[Fe/H]}}$ &   Linear  & $[-1,\ 0]$   &$[-2.5,\ -1.5]$&$[-1.5,\ -1]$\\
$\sigma_{\mathrm{[Fe/H]}}$ &   Linear  & $[0,\ 2]$   &$[0,\ 2]$&$[0,\ 2]$\\
$\mathcal{Q}$ &   Linear  & $[0,\ 1]$   &$\cdots$&$[0,\ 1]$\\
\bottomrule
\end{tabular}
}
\end{center}
\end{table}

We follow the framework in \citet{Necib_2019} to define the likelihood of each star $\mathcal{L}_{i}$, associated with a set of observables labeled as $O_{i}=(\boldsymbol{v}_{i},~{\feh}_{i})$ and different model parameters. We label the disk, halo and substructure components with flag of $j=d,h,s$, respectively. 

The likelihood of a star belonging to the disk component is
\begin{equation}
\label{eq1}
\begin{aligned}
&\mathcal{L}_{d}(O_{i}|\Theta)\\
&=\mathcal{N}(\boldsymbol{v}_{i}|\boldsymbol{\mu}^{d}, \boldsymbol{\Sigma}_{i}^{d})\mathcal{N}({\feh}_{i}|\mu^{d}_{{\feh}}, \sigma_{{\feh},i}^{d}) \ .
\end{aligned}
\end{equation}
Similarly, the likelihood function of the halo component is
\begin{equation}
\label{eq2}
\begin{aligned}
&\mathcal{L}_{h}(O_{i}|\Theta)\\
&=\mathcal{N}(\boldsymbol{v}_{i}|\boldsymbol{\mu}^{h}, \boldsymbol{\Sigma}_{i}^{h})\mathcal{N}({\feh}_{i}|\mu^{h}_{{\feh}}, \sigma_{{\feh},i}^{h}) \ ,
\end{aligned}
\end{equation}
where $\Theta$ represents the parameter set and $\mathcal{N}$ denotes the Gaussian distribution (3 dimensional Gaussian distribution for the velocity). $\Theta$ contains the free parameters of the Gaussian distributions for both the velocity and metallicity. ${\boldsymbol{\mu}^{j}}$ is the mean velocity ($\mu_{r,j},\mu_{\theta,j},\mu_{\phi,j}$), while $\boldsymbol{\Sigma}^{j}$ is the covariance matrix determined by the velocity dispersion $\sigma_{r,\theta,\phi}$ and the correlation coefficients $\rho_{r\theta,r\phi,\theta\phi}$. The correlation coefficients are found to be very small \citep[e.g.][]{Evans_2016,Necib_2019,Wu_2022}. In observations, the dispersion of metallicity and the velocity covariance matrix is a combination of the true value and the measurement errors, i.e. $\boldsymbol{\Sigma}_{\rm{obs}}=\boldsymbol{\Sigma}_{\rm{true}}+\boldsymbol{\Sigma}_{\rm{err}}, \sigma_{{\feh},\rm{obs}}=\sqrt{\sigma_{{\feh},\rm{true}}^{2}+\sigma_{{\feh},\rm{err}}^{2}}$, varying for each star. Considering the mean and covariance matrix of velocity, and the mean and dispersion of metallicity, the models of disk and halo components each contain 11 free parameters.

\begin{figure*}[tp]
\centering
\includegraphics[width=\textwidth]{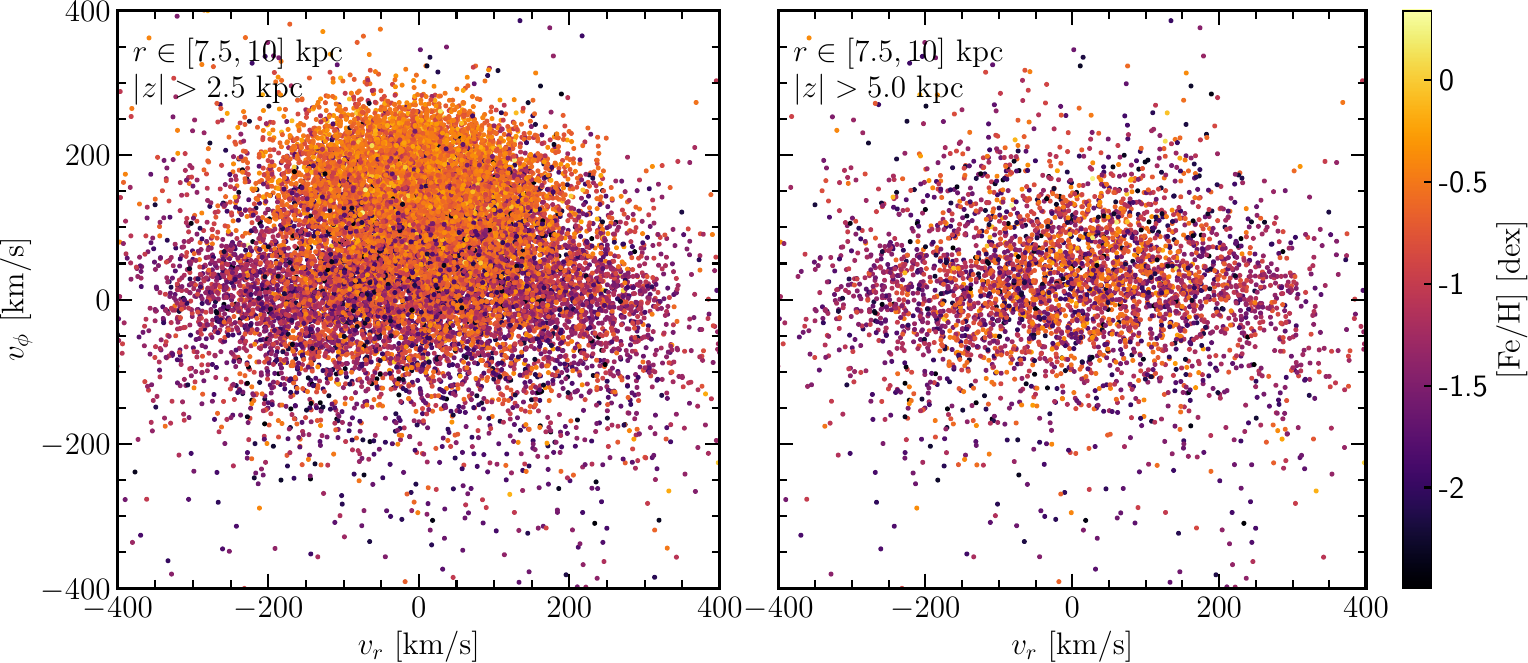}
\caption{Radial velocity ($v_{r}$) vs. azimuthal velocity ($v_{\phi}$) color-coded by \feh\ for the stellar samples within $r\in[7.5,10]$~\kpc\ $\&\ |z|>2.5$~{\kpc} (left) and $r\in[7.5,10]$~\kpc\ $\&\ |z|>5.0$~{\kpc} (right).}
\label{Fig.VR_VP}
\end{figure*}

\citet{Belokurov_2018b} and \citet{Necib_2019} demonstrated that the velocity of the substructure component resulted from the GES is highly radially anisotropic. Thus it can be described by a bimodal Gaussian distribution with a likelihood of 
\begin{equation}
\label{eq3}
\begin{aligned}
\mathcal{L}_{s}(O_{i}|\Theta)=&\frac{1}{2}[\mathcal{N}(\boldsymbol{v}_{i}|\boldsymbol{\mu}^{\tilde{s}}, \boldsymbol{\Sigma}_{i}^{s})+\mathcal{N}(\boldsymbol{v}_{i}|\boldsymbol{\mu}^{s}, \boldsymbol{\Sigma}_{i}^{s})]\\
&\times\mathcal{N}({\feh}_{i}|\mu^{s}_{{\feh}}, \sigma_{{\feh},i}^{s}) \ ,
\end{aligned}
\end{equation}
where $\boldsymbol{\mu}^{s}\!=\!(\mu_{r,s},\mu_{\theta,s},\mu_{\phi,s})$ and $\boldsymbol{\mu}^{\tilde{s}}\!=\!(-\mu_{r,s},\mu_{\theta,s},\mu_{\phi,s})$. This model resembles a single Gaussian distribution in $v_{r}$ when $\mu_{r,s}$ is small and $\sigma_{r,s}$ is large, in which case it results in a strong degeneracy between the halo and the substructure components. The model of the substructure component gives another 11 free parameters.

Thus, the likelihood for a sample containing $N$ stars is defined as 
\begin{equation}
\label{eq4}
\mathcal{L}(\{O_{i}\}|\Theta)=\prod_{i=1}^{N}\sum_{j=d,h,s}\mathcal{Q}_{j}\mathcal{L}_{j}(O_{i}|\Theta) \ ,
\end{equation}
where $\mathcal{Q}_{j}$ denotes the probability that the star belongs to the \textit{j}-th component with an normalization constraint of $\mathcal{Q}_{h}=1-\mathcal{Q}_{d}-\mathcal{Q}_{s}$. $\mathcal{Q}_{j}$ also represents the integrated fractional area of any marginalized 1D distribution for that component. With the two fraction parameters of $\mathcal{Q}_{d}$ and $\mathcal{Q}_{s}$, we have 35 free parameters in total. 

We apply the \textbf{emcee} \citep{Foreman_Mackey_2013} package to perform the Markov Chain Monte Carlo (MCMC) Bayesian estimation to find the posterior distributions of all 35 free parameters. We utilize 250 walkers and 5000 steps as burn-in, followed by another 5000 steps to obtain the posterior distributions of parameters. The flat priors of each parameter are listed in Table~\ref{table1}, similar to \cite{Necib_2019}. Due to the two obvious peaks at $\feh \sim-1.2$ and $\feh \sim-2.0$ in the metallicity distribution and the values obtained in previous works \citep{Necib_2019, Wu_2022}, we utilize narrower prior ranges for the mean metallicities. We tested that the results are nearly the same if we adopt the same prior as \citet{Necib_2019}. We adopt the set of parameters with the maximum likelihood as the best-fit result and the 2.1\%, 50\%, and 97.9\% percentiles of the 1D marginalized posterior distributions to denote the median and $\sim2\sigma$ uncertainties.

\section{The Stellar Distribution in the Chemodynamical Space}
\label{Sec3: Stellar Distribution in the Chemodynamical Space}

As illustrated by \citet{S_B_2012} and \citet{Necib_2019}, below $|z|\sim2.5$~{\kpc}, the azimuthal velocity distribution of the disk cannot be fitted with a single Gaussian distribution due to the contribution from the thin disk. Thus, in this Section, we focus on the analyses of the stars within $r\in[7.5, 10.0]$~kpc and $|z|>$ 2.5~kpc. The velocity distribution of $v_{r}$ versus $v_{\phi}$ of this sample is displayed in the left panel of Fig. \ref{Fig.VR_VP}. The clump with a high metallicity and an azimuthal rotation of $\sim 200$ {\kms} is the disk component, which is the dominant component in this sample. The elongated ``sausage'' structure in $v_{r}$ direction with a lower metallicity is the substructure component contributed by the GES. It becomes the dominant component when a larger vertical cut of $|z|>$ 5~kpc is applied (right panel of Fig. \ref{Fig.VR_VP}) where the disk component disappears. In order to extend the GMM fitting method to the Galactic plane, we apply an additional angular momentum ($L_{z}$) cut to reduce the contribution of the disk component in the later Section~\ref{Sec4: Angular Momentum Constrain}.

\subsection{The Solar Neighbourhood}
\label{The Solar Neighbourhood}

 \begin{figure*}[htbp]
\centering
\includegraphics[width=0.75\textwidth]{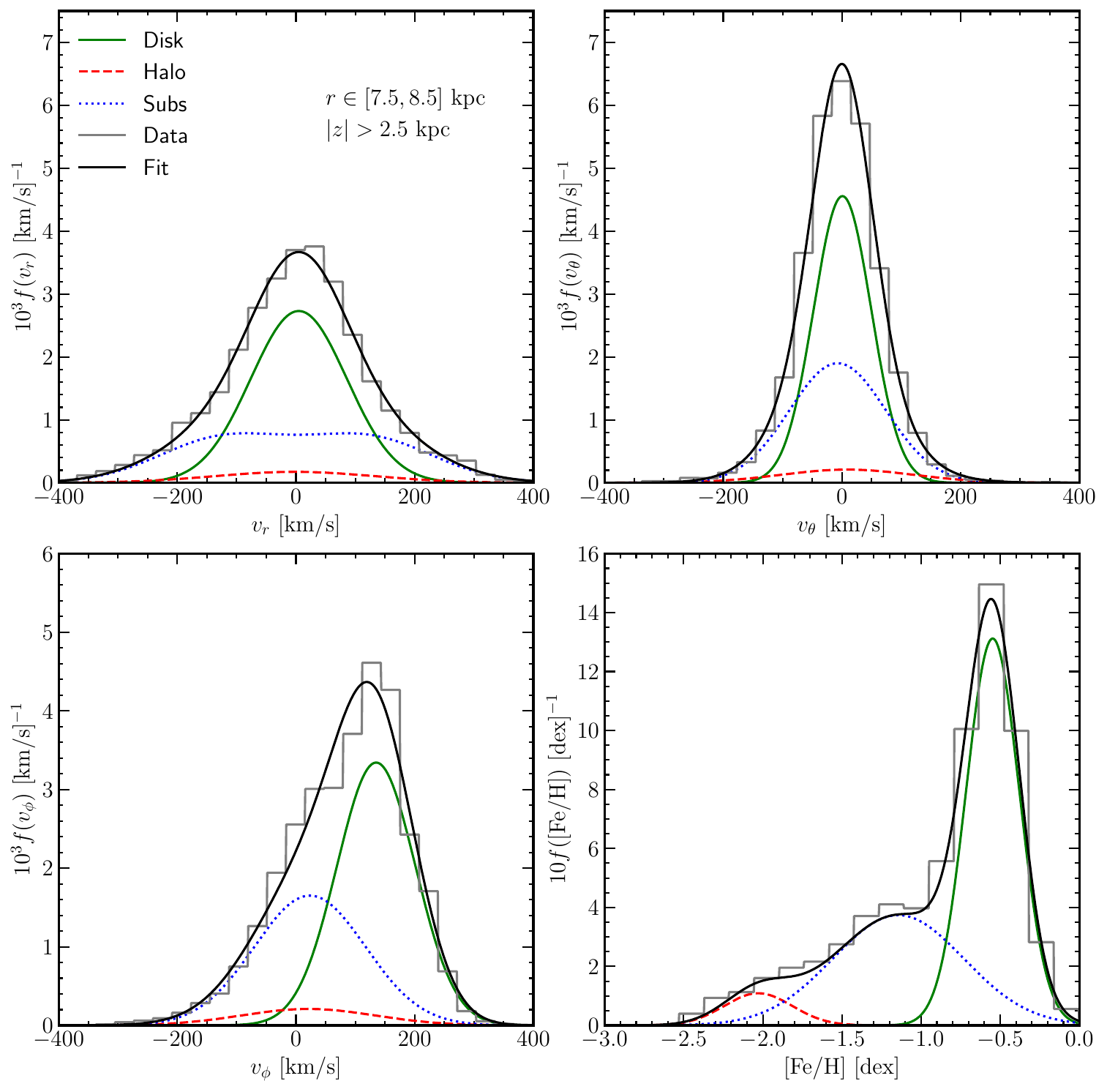}
\caption{The best-fit chemodynamical distribution ($v_{r},v_{\theta},v_{\phi},\feh$) of the sample in the Solar neighbourhood with $r\in[7.5,8.5]~${\kpc} \& $|z|>2.5~${\kpc}. The gray histogram is the data, while the colored lines are from the model which represent the disk, substructure and halo components as green, blue and red lines, respectively.}
\label{SN}
\end{figure*}

As we aim to obtain the local DM velocity distribution, we select a sample in the Solar neighbourhood with a radial range of $r\in[7.5,8.5]~${\kpc} and a vertical cut of $|z|>2.5~${\kpc}. We apply the GMM method and the MCMC technique to separate the three components. The results are shown in Fig.~\ref{SN}, and the posterior distributions of the free parameters are displayed in the Appendix for each component separately. Our best-fit model well fits the chemodynamical distributions of the sample. The disk component is the most dominant, while the halo component contributes the least. The fractions of the three components are $\mathcal{Q}_{d}=0.55^{+0.02}_{-0.02}$, $\mathcal{Q}_{s}=0.39^{+0.02}_{-0.02}$ and $\mathcal{Q}_{h}=0.06^{+0.02}_{-0.01}$, for the disk, substructure and halo components, respectively. The relative fraction between the substructure and halo components is $\mathcal{Q}_{s}/(\mathcal{Q}_{h}+\mathcal{Q}_{s})=0.87^{+0.03}_{-0.04}$, which means that the substructure contributes $\sim 87\%$ in all local halo stars. It should be noted that there are potential impacts of the selection effect of the spectroscopic survey (sample incompleteness compared to the photometric survey) on this fraction. \citet{Wu_2022} shows that the LAMOST selection function mainly affects the GES fraction $\mathcal{Q}_{s}/(\mathcal{Q}_{h}+\mathcal{Q}_{s})$ for fainter stars in the very outer halo ($r<30$~{\kpc}). For brighter stars in the Solar neighbourhood, the impact of selection function is negligible as shown in their Fig.~24. The selection effect are not taken into consideration in this study since our K giants are within $r<10$~{\kpc}.


The disk component is relatively metal-rich with a mean metallicity of $\mu_{\feh}=-0.55_{-0.01}^{+0.01}$ dex and a dispersion of $\sigma_{\feh}=0.17_{-0.01}^{+0.01}$ dex. Its motion is dominated by the azimuthal rotation with $\mu_{\phi}=134.73_{-2.87}^{+2.74}$ \kms\ and $\sigma_{\phi}=65.89_{-2.17}^{+2.12}$ \kms. The substructure component is radially anisotropic, showing a bimodal Gaussian distribution in $v_{r}$ with $\mu_{r}=120.09_{-6.61}^{+6.44}$ \kms\ and $\sigma_{r}=108.52_{-4.53}^{+5.20}$ \kms. As we adopt the right-handed Galactocentric frame, $v_{\phi}>0$ means a prograde rotation. Thus, for the substructure component, $\mu_{\phi}=22.59_{-4.61}^{+4.74}$ \kms\ and $\sigma_{\phi}=93.43_{-3.29}^{+3.20}$ {\kms} indicates a slightly prograde orbit as referred in some previous works about the debris flow of the massive radial merger event GES~\citep{Belokurov_2018b,Necib_2019,Wu_2022}, whereas other works show a retrograde rotation \citep[e.g.][]{Helmi_2018,Koppelman_2020a,Mackereth_2019}. Our sample predicts a more metal-rich substructure component with $\mu_{\feh}=-1.15_{-0.04}^{+0.04}$ dex than the previous result of $\mu_{\feh}=-1.39$ dex \citep{Necib_2019}. The difference could result from the degeneracy between the substructure and halo components. The halo component consists of the accreted stars from the oldest mergers, which thus becomes relaxed and isotropic with almost zero mean velocities and large dispersions,~i.e. $(\mu_{r},\mu_{\theta},\mu_{\phi})=(-4.59_{-18.19}^{+16.53},8.02_{-13.30}^{+13.74},19.82_{-14.06}^{+14.19})$~{\kms}\,and\,$(\sigma_{r},\sigma_{\theta},\\\sigma_{\phi})=(128.47_{-12.97}^{+12.85},113.62_{-9.52}^{+11.59},112.02_{-9.69}^{+11.00})$~\kms. Note that the halo component also shows a slightly prograde azimuthal rotation similar to the substructure component. The metallicity of the halo component is more metal-poor than many previous works with $\mu_{\feh}=-2.04_{-0.06}^{+0.07}$ dex and $\sigma_{\feh}=0.21_{-0.04}^{+0.04}$ dex.


All of the velocity correlation coefficients for each component are small as expected: for the disk component $(\rho_{r\theta},\rho_{r\phi},\rho_{\theta\phi})=(0.16_{-0.04}^{+0.04},-0.09_{-0.04}^{+0.04},0.02_{-0.04}^{+0.04})$; for the substructure component $(\rho_{r\theta},\rho_{r\phi},\rho_{\theta\phi})\!\!=\!\!(0.15_{-0.05}^{+0.05},-0.08_{-0.06}^{+0.06},0.06_{-0.04}^{+0.04})$; for the halo component $(\rho_{r\theta},\rho_{r\phi},\rho_{\theta\phi})=(-0.07_{-0.12}^{+0.12},0.03_{-0.13}^{+0.16},-0.08_{-0.12}^{+0.12})$.

\begin{figure*}[htbp]
\centering
\includegraphics[width=\linewidth]{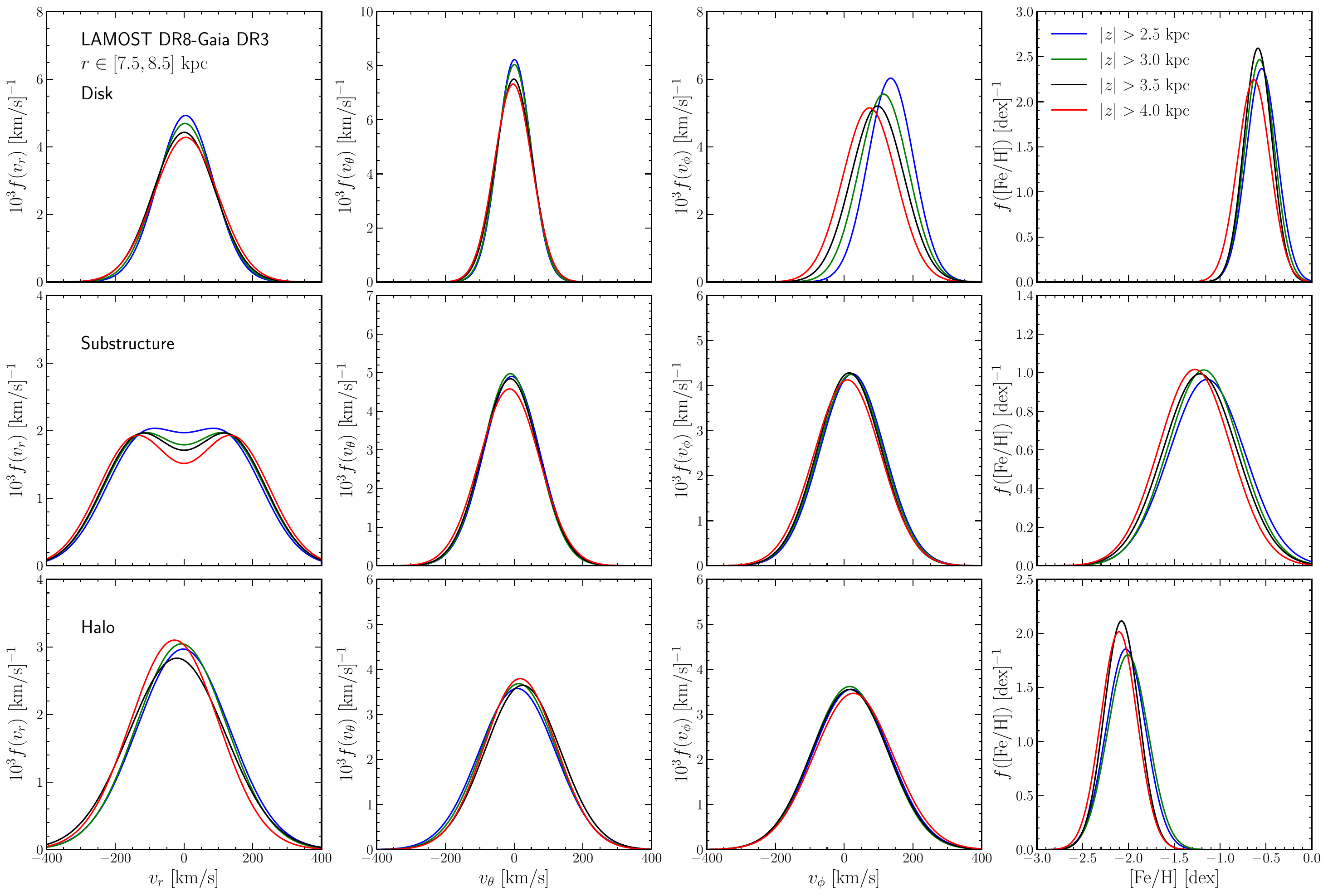}
\caption{Best-fit chemodynamical distributions for samples with different vertical height cuts. We vary the vertical height constraints from $|z|>2.5$ to $4.0$~\kpc, while the radial range is the same i.e. $r\in[7.5,8.5]$~\kpc. The columns from the left to the right are distributions of $v_{r},v_{\theta},v_{\phi},\rm\feh$, respectively. The rows from the top to the bottom are for the disk, substructure and halo components, respectively. The distributions are nearly invariant as $z_{\rm{cut}}$, except for the $v_{\phi}$ distribution of the disk component, which shifts to a lower velocity as the sample goes far away from the midplane.}
\label{vertical_variation_distribution}
\end{figure*}

\subsection{Spatial Variation}
\label{Spatial Variation}
We applied a vertical constraint of $|z|>2.5~${\kpc} to avoid the strong dominance of the disk component in lower vertical region. In this Section, we explore the spatial variation of our results by applying different radial and vertical cuts to the samples to check if the results could be extrapolated to the disk midplane.

\subsubsection{Vertical Variation}
\label{Vertical Variation}

\begin{figure*}[htp]
\centering
\includegraphics[width=\linewidth]{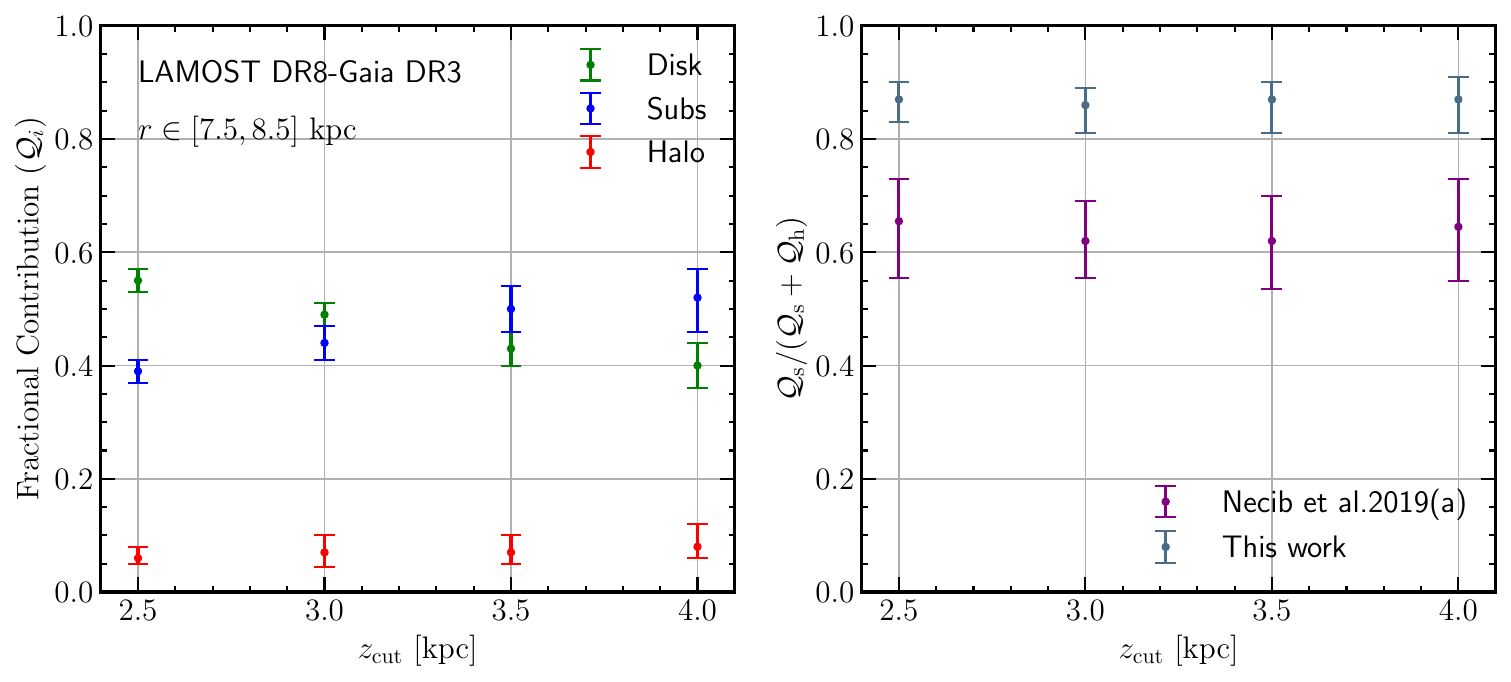}
\caption{Left: fractional contribution of different components as a function of vertical height cut for samples within $r\in[7.5,8.5]~\rm{\kpc}$ \& $|z|>z_{\rm cut}$. The green, blue and red dots are for the disk, substructure and halo components, respectively. Right: fraction of the substructure relative to the nondisk components. The error-bars represent the 2.1\%, 50\% and 97.1\% values of the MCMC results. }
\label{vertical_variation}
\end{figure*}

To inspect the vertical variation, we select stars within $r\in[7.5,8.5]$~\kpc, and apply different vertical cuts of $|z|>z_{\rm{cut}}$ , varying $z_{\rm{cut}}$ from 2.5 to 4 \kpc. Fig.~\ref{vertical_variation_distribution} displays the best-fit models for the different vertical samples, with different rows representing the different components and different columns for the chemodynamical distributions. Most distributions of the three components do not vary much with the vertical cuts, except for a decreasing azimuthal rotation with an increasing $z_{\rm{cut}}$, which is expected. The radial velocity of the substructure shows slightly stronger bimodal distribution as $z_{\rm{cut}}$ increases. Fig.~\ref{vertical_variation} shows the fractional contributions of the different components. The fraction of the disk decreases with $z_{\rm{cut}}$ as expected, while the contributions of the substructure and halo components both increase. Nevertheless, the relative fraction of substructure component to the nondisk components are nearly unchanged with $\mathcal{Q}_{s}/(\mathcal{Q}_{s}+\mathcal{Q}_{h})\sim0.85$.


\subsubsection{Radial Variation}
\label{Radial Variation}
\begin{figure*}[htbp]
\centering
\includegraphics[width=\linewidth]{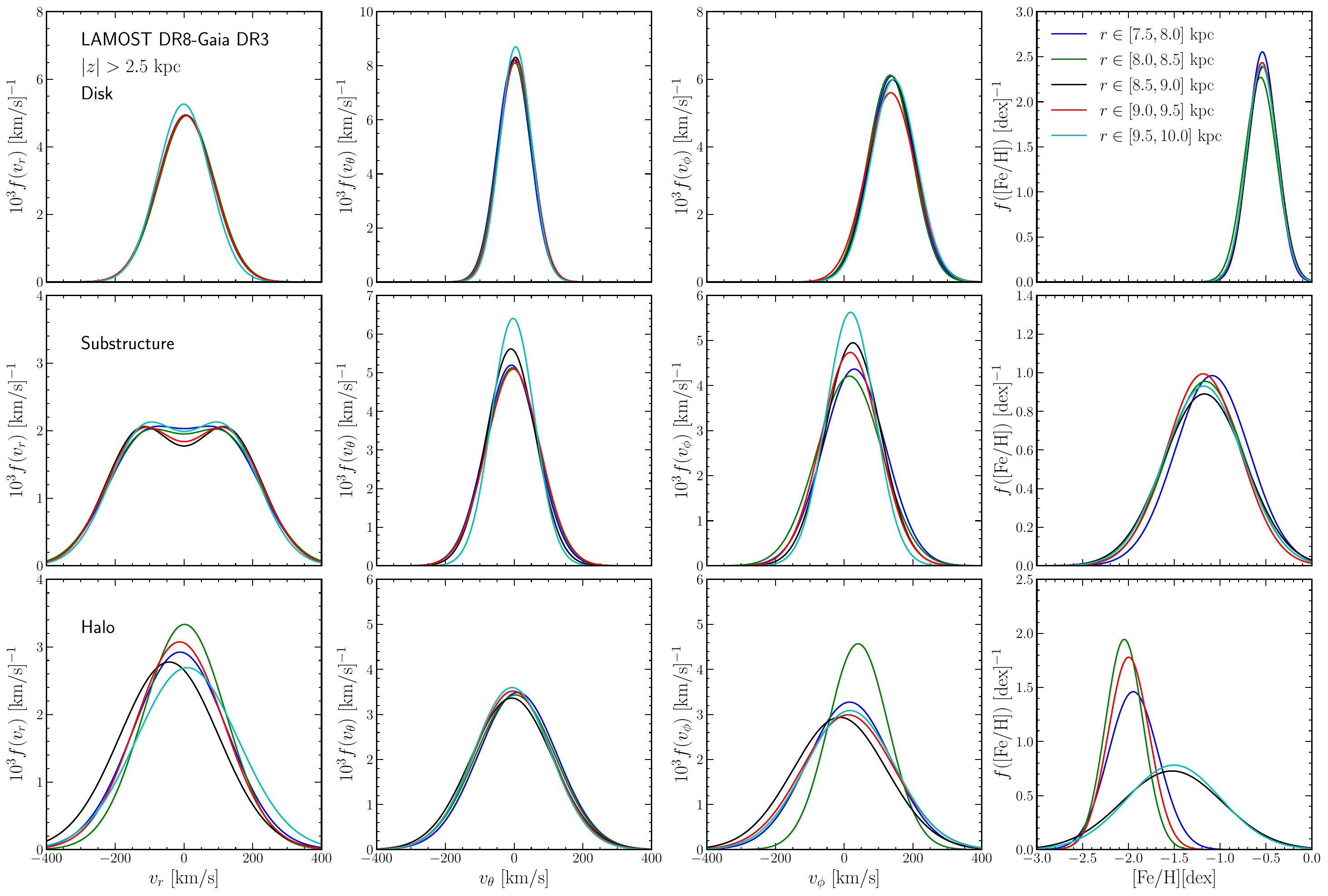}
\caption{Same as Fig.~\ref{vertical_variation} but for samples with the same vertical cut ($|z|>2.5$~\kpc) and different radial ranges.}
\label{radial_variation_distribution}
\end{figure*}

\begin{figure*}[tp]
\centering
\includegraphics[width=\linewidth]{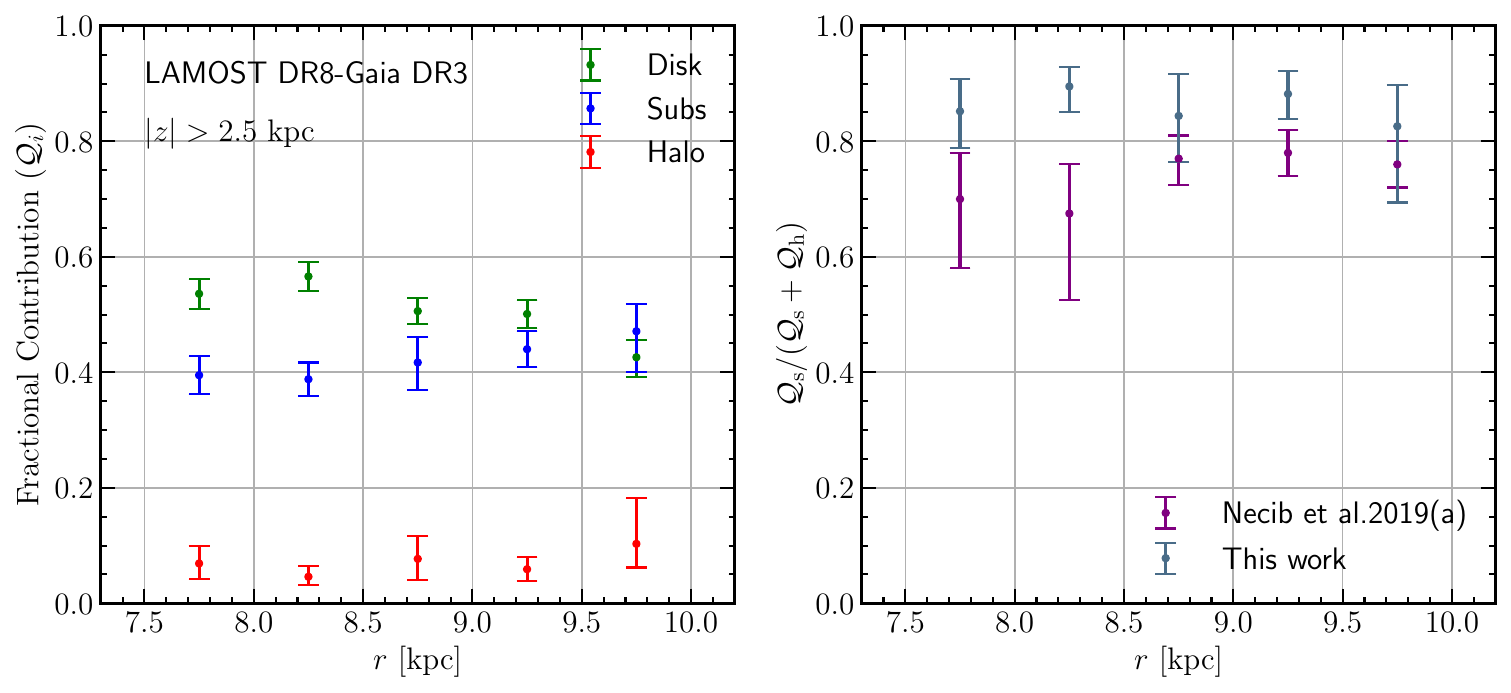}
\caption{Same as Fig.~\ref{vertical_variation} but for samples with a same vertical cut ($|z|>2.5$~\kpc) and different radial range. }
\label{radial_variation}
\end{figure*}

Similarly, we explore the radial variation of the results by selecting subsamples with the same vertical cut of $|z|>2.5$~{\kpc} and different radial cuts, varying from 7.5 to 10.0 {\kpc} with a bin size of 0.5 kpc. The results are shown in Figs.~\ref{radial_variation_distribution} and \ref{radial_variation}. As shown in Fig.~\ref{radial_variation_distribution}, the chemodynamical distributions of the disk and substructure are invariant for various radial ranges, while the halo component shows some variation, especially in the metallicity. In the range of $r\in[8.5,9.0]$ \&\ $[9.5,10]$~{\kpc}, the mean metallicity is much higher with $\mu_{\feh}\sim -1.5$ dex (which is the upper limit of the prior) and  a larger dispersion than the other three radial bins. This metallicity distribution is more similar to that of the substructure component, which could be due to the strong degeneracy between the two components. In addition, the halo fraction is always less than 10\% in our samples, which means that its chemodynamical distributions could be dominated and influenced by the other two components. We also discuss fitting the data with a two-component model without the halo component to inspect the degeneracy in Section.~\ref{Sec6.2: Degeneracy between Substructure and Halo}. Nevertheless, as shown in the right panel of Fig.~\ref{radial_variation}, the relative fraction of substructure and halo still varies within 5\% around $\mathcal{Q}_{s}/(\mathcal{Q}_{s}+\mathcal{Q}_{h})\sim0.85$.


\subsection{Parameter Degeneracy}
\label{Parameter Degeneracy}
There are 35 free parameters in total for the three components in the GMM. It is important to inspect the degeneracy between the different parameters and different components. The posterior distributions of these parameters except for the fractions ($\mathcal{Q}_{h},\ \mathcal{Q}_{d},\ \mathcal{Q}_{s}$) are illustrated in the Appendix, for each component separately. There is strong degeneracy between the mean and dispersion of the metallicity for all three components. More sepecifically, $\mu_{\feh}$ and $\sigma_{\feh}$ of the disk and substructure components show strong negative correlations, while they show a positive correlation for the halo component. We also find negative correlations between the mean and dispersion of the azimuthal rotation of the disk component ($\mu_{\phi,d}$ vs. $\sigma_{\phi,d}$) and of the radial motion of the substructure component ($\mu_{r,s}$ vs. $\sigma_{r,s}$). 

\begin{figure*}[t]
\centering
\includegraphics[width=\linewidth]{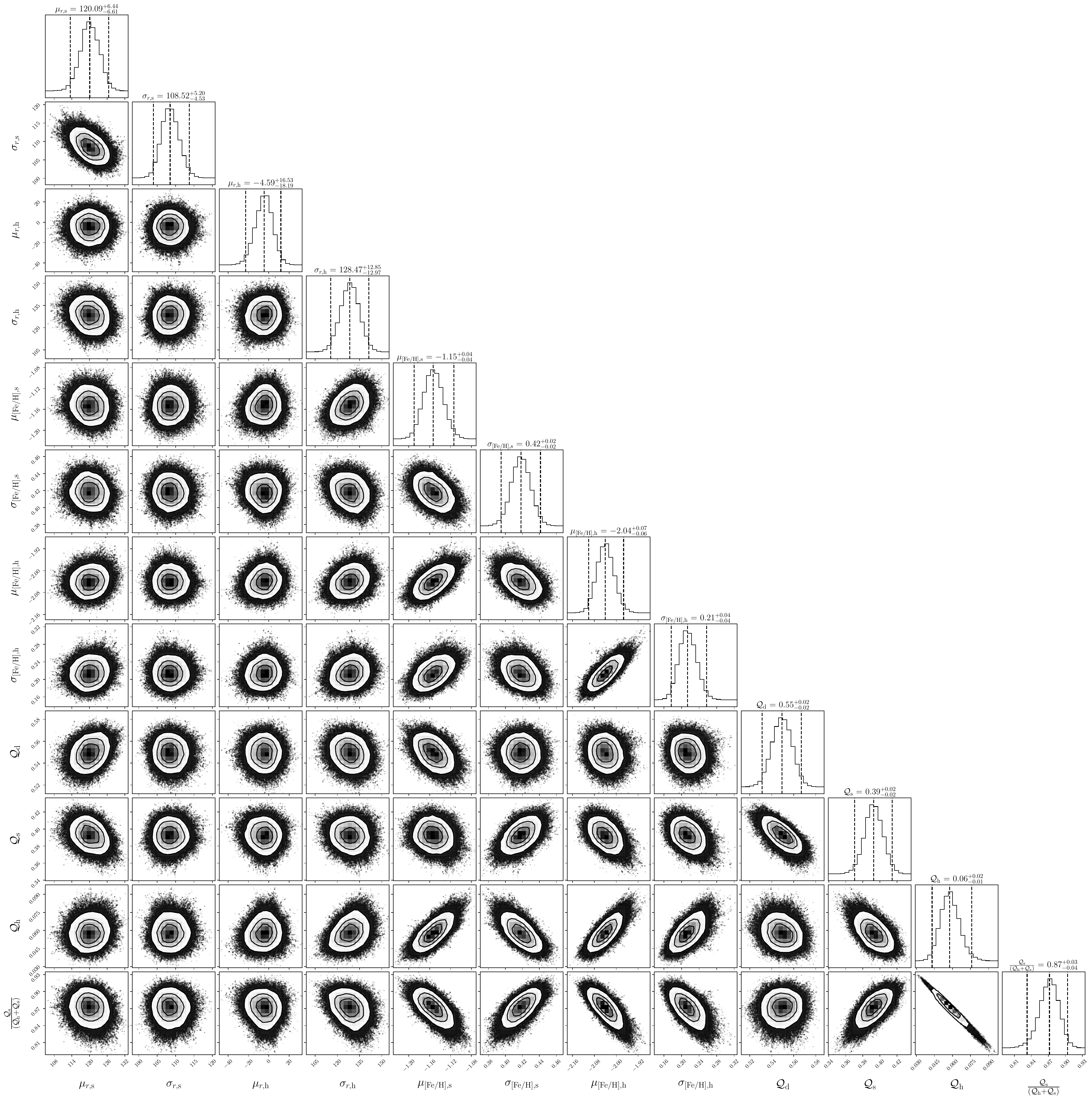}
\caption{Posterior distributions of possible degenerate model parameters from the MCMC results of the sample in the region $r\in[7.5,8.5]$~{\kpc} \& $|z|>2.5$~{\kpc}. The dashed lines of each 1D distribution represent the 2.1\%,50\%, and 97.9\% percentiles.}
\label{corner_deg}
\end{figure*}

As the disk component is significantly different from the other two components, the latter two could have a strong degeneracy, especially between the radial motion and metallicity. Thus, we show the posterior distributions of the parameters of $\mu_{r}$, $\sigma_{r}$, $\mu_{\feh}$, $\sigma_{\feh}$ for the substructure and halo components, as well as the fraction parameters in Fig.~\ref{corner_deg}. Many correlations exist. (1) As expected both the fractions of the disk ($\mathcal{Q}_{d}$) and the halo ($\mathcal{Q}_{h}$) components show negative correlations with the fraction of the substructure ($\mathcal{Q}_{s}$) component, though these two show no correlation. (2) The fraction of the halo component $\mathcal{Q}_{h}$ shows strong correlations with the parameters of the metallicity of the substructure and halo components. This is due mainly to that the halo component in our sample contributes quite less with $\mathcal{Q}_{h} < 0.1$. Its lower metallicity leads it to emerge in the poor end of the metallicity distribution in our sample. (3) The mean radial velocity of the substructure $\mu_{r,s}$ displays slight correlations with the fraction parameters $\mathcal{Q}_{d}$ and $\mathcal{Q}_{s}$. (4) The parameters of metallicity of the substructure and halo components show strong correlations with each other. (5) The relative fraction of the substructure in the nondisk components $\mathcal{Q}_{s}/(\mathcal{Q}_{s}+\mathcal{Q}_{h})$ is strongly affected by the metallicity parameters. Though its value is not affected by the fraction of disk component $\mathcal{Q}_{d}$, its determination is strongly influenced by the poor end (\feh \,$< -1$) of the metallicity distribution of the sample. It reveals the importance of the metallicity distribution of the sample for separating these two components.

\section{Angular Momentum Constraints}
\label{Sec4: Angular Momentum Constrain}

Closer to the midplane, the disk component is more dominant and the identification of the substructure and halo components is more difficult. In order to reduce the contribution and influence of the disk component, we have tried many methods to remove the disk stars, e.g. the vertical height cut (Section \ref{Sec3: Stellar Distribution in the Chemodynamical Space}), angular momentum cut, metallicity cut and Toomre diagram cut. In this section, we apply an additional angular momentum ($L_{z}$) constraint to our samples, which can efficiently reduce the disc dominance for regions closer to the disk midplane while keeping the applicability of the GMM. The other two methods and their shortcomings are discussed in more details in Section~\ref{Sec6.1: Methods of Decreasing the Contribution of Disk Stars}. As the substructure and halo components have slightly prograde rotation with quite small mean rotation of $\mu_{\phi}\sim20$~\kms, we utilize the angular momentum cut of $|L_{z}|<500$~\kpc\,\kms \citep{Feuillet_2020,Belokurov_2022}, corresponding to $v_{\phi}\sim62.5$~\kms\ at $R=8$~\kpc. 

Before moving closer to the midplane, we apply the angular momentum cut to the sample within $r\in[7.5,8.5]~${\kpc} \& $|z|>2.5~${\kpc}, and test the performance of the GMM. The best-fit chemodynamical distributions of this constrained sample are shown in Fig.~\ref{Lzcut}. Fig.~\ref{Lzcutcom} shows the comparison of the best-fit models with (dotted lines) and without (solid lines) the angular momentum cut, with the 2.1\%, 50\% and 97.9\% percentiles of posterior distributions of parameters in the GMM listed in Tables~\ref{table3}-\ref{table5} for the three components, respectively.

The $v_{\phi}$ distribution of all three components have smaller $\sigma_{\phi}$ as expected. The prograde rotation of the nondisk components still remains, though the mean rotation $\mu_{\phi}$ decreases from $\sim20$~\kms\ to $\sim4$~\kms\ and $\sim14$~\kms\ for the substructure and halo components, respectively, as shown in Tables~\ref{table3} and \ref{table4}. For all three components, the medians of the best-fit distributions of $v_{r}$, $v_{\theta}$, \feh\ are similar for the results with and without the angular momentum constraint, as shown in Fig.~\ref{Lzcutcom} and Tables~\ref{table3}-\ref{table5}. The angular momentum constraint significantly decreases the contribution of the disk component and the substructure becomes dominant, with component fractions of $\mathcal{Q}_{d}=0.36^{+0.04}_{-0.03}$, $\mathcal{Q}_{s}=0.55^{+0.05}_{-0.05}$ and $\mathcal{Q}_{h}=0.09^{+0.04}_{-0.03}$. The resultant relative fraction of the substructure and halo components is $\mathcal{Q}_{s}/(\mathcal{Q}_{s}+\mathcal{Q}_{h})=0.86^{+0.05}_{-0.06}$. However, as the $v_{\phi}$ distributions of these two components have been changed, the relative fractions of stars removed could be different. In order to roughly quantify and correct such a bias, we apply the $L_{z}$ constraint to the MCMC best-fit distribution obtained for the sample without the $L_{z}$ constraint, i.e. the results shown in Fig.~\ref{SN}. We calculate the probability of a star belonging to each component, and select stars with highest probability belonging to the substructure or halo components to obtain their spatial distribution. We then sample $\sim10^6$ particles for each star based on the best-fit model from GMM to get the velocity distribution. Afterwards, we apply the $L_{z}$ constraint to the pseudo sample and estimate the removed fraction due to the $L_{z}$ cut for the substructure or halo components, respectively. The removed fraction is applied to the results of the sample in this section to roughly correct the influence of the $L_{z}$ constraint. After correction, $\mathcal{Q}_{s}/(\mathcal{Q}_{s}+\mathcal{Q}_{h})\sim0.8$, which is consistent with the results in Section~\ref{The Solar Neighbourhood}.

Then we inspect the variation of the results with the vertical cut $z_{\rm cut}$ for samples with the $L_{z}$ cut and additionally study two vertical regions closer to the midplane. We select stars within $r\in[7.5,8.5]$~\kpc\ \& $|L_{z}|<500$~\kpc\,\kms, and apply different vertical cuts of $|z|>z_{\rm{cut}}$, varying $z_{\rm{cut}}$ from 0.5 to 4~\kpc, including two smaller $z_{\rm cut}$ of 0.5 and 1.5 \kpc\footnote{With the angular momentum constraint, the samples are almost the same for applying the vertical cuts $z_{\rm{cut}}=0$ or $0.5$~\kpc\ in this radial range. Therefore the case of $z_{\rm{cut}}=0$~\kpc\ is almost the same as $z_{\rm{cut}}=0.5$~\kpc.}. Fig.~\ref{veritical_variation_distribution_Lz} shows the vertical variation of the best-fit chemodynamical distributions for the $L_{z}$-cut samples similar to Fig.~\ref{vertical_variation_distribution}. The angular momentum constraint cuts mainly the high rotation part of the disk component, which is more significant at lower vertical height. The remaining stars in samples with different $z_{\rm{cut}}$ show almost invariant chemodynamical distributions for all three components, even for the two samples closer to the midplane. As shown in Tabel.~\ref{table4}, for larger vertical cuts e.g. $z_{\rm{cut}}=3.5,4.0$~\kpc, the parameters of the halo component have larger uncertainties when the $L_{z}$ constraint applied, which could be caused by the degeneracy between the substructure and halo components. The radial velocity distribution of the substructure component also shows slightly stronger bimodal distribution with increasing $z_{\rm{cut}}$. As the velocity correlation $|\rho_{r\phi}|$ \& $|\rho_{\theta\phi}|\lesssim0.1$, we can regard $v_{\phi}$ to be independent on $v_{r}$ \& $v_{\theta}$ for each component. Therefore, though the $v_{\phi}$ distribution has been significantly changed by the $L_{z}$ cut, the near invariance of the results with $z_{\rm{cut}}$ means that we can safely extrapolate our results to the Galactic plane.


Fig.~\ref{spatial_variation} shows the comparison of fractions of the substructure component relative to the nondisk components for results with (red) and without (blue) the angular momentum constraint. The results from \citet{Necib_2019} are also shown as the purple dots. This fraction is consistent for samples with different vertical and radial cuts, and also for samples with or without the angular momentum constraint. It demonstrates that for our sample $\mathcal{Q}_{s}/(\mathcal{Q}_{s}+\mathcal{Q}_{h})=0.8-0.85$, which is slightly higher than that obtained in \citet{Necib_2019} of $\mathcal{Q}_{s}/(\mathcal{Q}_{s}+\mathcal{Q}_{h})=0.6-0.8$. Note that the actual $\mathcal{Q}_{s}/(\mathcal{Q}_{s}+\mathcal{Q}_{h})$ ratio that \citet{Necib_2019} adopted to deduce the DM velocity distribution is $76\%$, a value computed using all stars with $d_{\odot}<4$~\kpc~\&~$|z|>2.5$~\kpc.


\section{Dark Matter Velocity Distribution and Detection Limits}
\label{Sec5: Dark Matter Distribution and Detection Limits}

In this section we construct the local DM velocity distribution for the dual DM model (i.e. the substructure and halo components), and estimate its influence on the DM direct detection limits compared to the SHM. In Section~\ref{DM Distribution} we present the conversion of the relative fraction of the substructure component in the observed nondisk stellar components to the fraction of the substructure component in the total DM, based on which we construct the local DM velocity distribution. In Section~\ref{Detection Limits} we show the impacts on the direct detection limits where the spin-independent nuclear scattering is considered.

\subsection{DM Velocity Distribution}
\label{DM Distribution}

In the SHM the DM halo is spherical and isotropic following the Maxwell-Boltzmann distribution in the Galactocentric frame with the circular velocity of the local standard of rest (LSR) $v_{0}=238~${\kms} \citep{SR_2012}. The velocity distribution is truncated at the escape velocity $v_{esc}=544$~\kms\ at the Sun based on the work of \citet{Smith_2009} to construct finite extent halo which is also applied to the following dual halo model.

The dual halo consists of a DM substructure and a virialized SHM halo. In order to construct the local DM velocity distribution from the stellar sample, we follow the framework in \citet{Necib_2019b} from which we can reproduce their $c_{s}~\&~c_{h}$ in the following equation including the error ranges. The velocity distribution of the dual halo model is:
 \begin{equation}
 \label{eq5}
f_{\rm{dm}}(v)=\!N\!\left[\mathcal{Q}_{\ast,h}f_{h}(v)+\frac{c_{s}}{c_{h}}\mathcal{Q}_{\ast,s}f_{s}(v)\right] \ ,
 \end{equation}
 where $N$ is the normalization factor, $\mathcal{Q}_{\ast,j}=\frac{\mathcal{Q}_{j}}{\mathcal{Q}_{s}+\mathcal{Q}_{h}}\,(j = s,h)$, and $f_{j}(v)$ is the best-fit velocity distribution given by the MCMC results in Section~\ref{The Solar Neighbourhood}, based on the correlation between the accreted stars and DM. $\mathcal{Q}_{j}$ utilized in the following computation is also from Section.~\ref{The Solar Neighbourhood}. $c_{j}=\frac{M_{\rm{peak}}}{M_{\ast,\rm{total}}}$, in which $M_{\rm{peak}}$ and $M_{\ast,\rm{total}}$ are the peak halo mass and the total stellar mass of the progenitor galaxy for each DM component, respectively. However, $M_{\rm{peak}}$ and $M_{\ast,\rm{total}}$ of the progenitor galaxies for the substructure and halo components are not given by observations directly. Therefore we need to correlate $M_{\rm{peak}}$ and $M_{\ast,\rm{total}}$ with observables like \feh. We utilize the SMHM (stellar mass-halo mass) relation and the \feh-$M_{\ast,\rm{total}}$ (metallicity-stellar mass) relation to derive the correlation between $\left< \feh\right>$ and $c_{j}$ as the following.

First, we generate $M_{\ast,\rm{total}}$ from $M_{\rm{peak}}$ by a Monte Carlo procedure based on the SMHM relation. As illustrated in \citet{Behroozi_2013}, this relation shows dependence on the redshift $z$, and different trends for $M_{\rm peak}$ larger or smaller than a critical mass $M_{1}\sim10^{11.5}M_{\odot}$. The general formalism is as below:
\begin{equation}
\setlength{\abovedisplayskip}{2pt}
\setlength{\belowdisplayskip}{2pt}
\log(M_{\ast,\rm{total}})\!=\!\log_{10}(\epsilon M_{1})\!+\!f\!\!\left[\log_{10}\!\!\left(\frac{M_{\rm peak}}{M_{1}}\right)\!\!\right]\!-\!f(0) \ ,
\label{eq6}
\end{equation} 
where
\begin{equation}
\setlength{\abovedisplayskip}{2pt}
\setlength{\belowdisplayskip}{2pt}
\label{eq7}
f(x)=-\log_{10}(10^{-\alpha x}+1)+\delta\frac{(\log_{10}(1+\rm{exp}(\it x)))^{\gamma}}{1+\rm{exp}(10^{-\it x})} \ .
\end{equation}

For any given $M_{\rm peak}$, $M_{\ast,\rm{total}}$ is randomly selected from the lognormal distribution with a median given by Eq.~\ref{eq6} and a scatter $\sigma$. As illustrated in \citet{Garrison_2017a}, for $M_{\rm peak}>M_{1}$, free parameters in Eqs.~\ref{eq6} \&~\ref{eq7}, i.e. $\epsilon,\alpha,\delta,\gamma$ and the scatter $\sigma$, are fixed to the values given in \citet{Behroozi_2013}. While for $M_{\rm peak}\leq M_{1}$, $\alpha\ \&\ \sigma$ are determined by the factor $\nu$ as:
\begin{figure*}[t]
\centering
\includegraphics[width=0.75\textwidth]{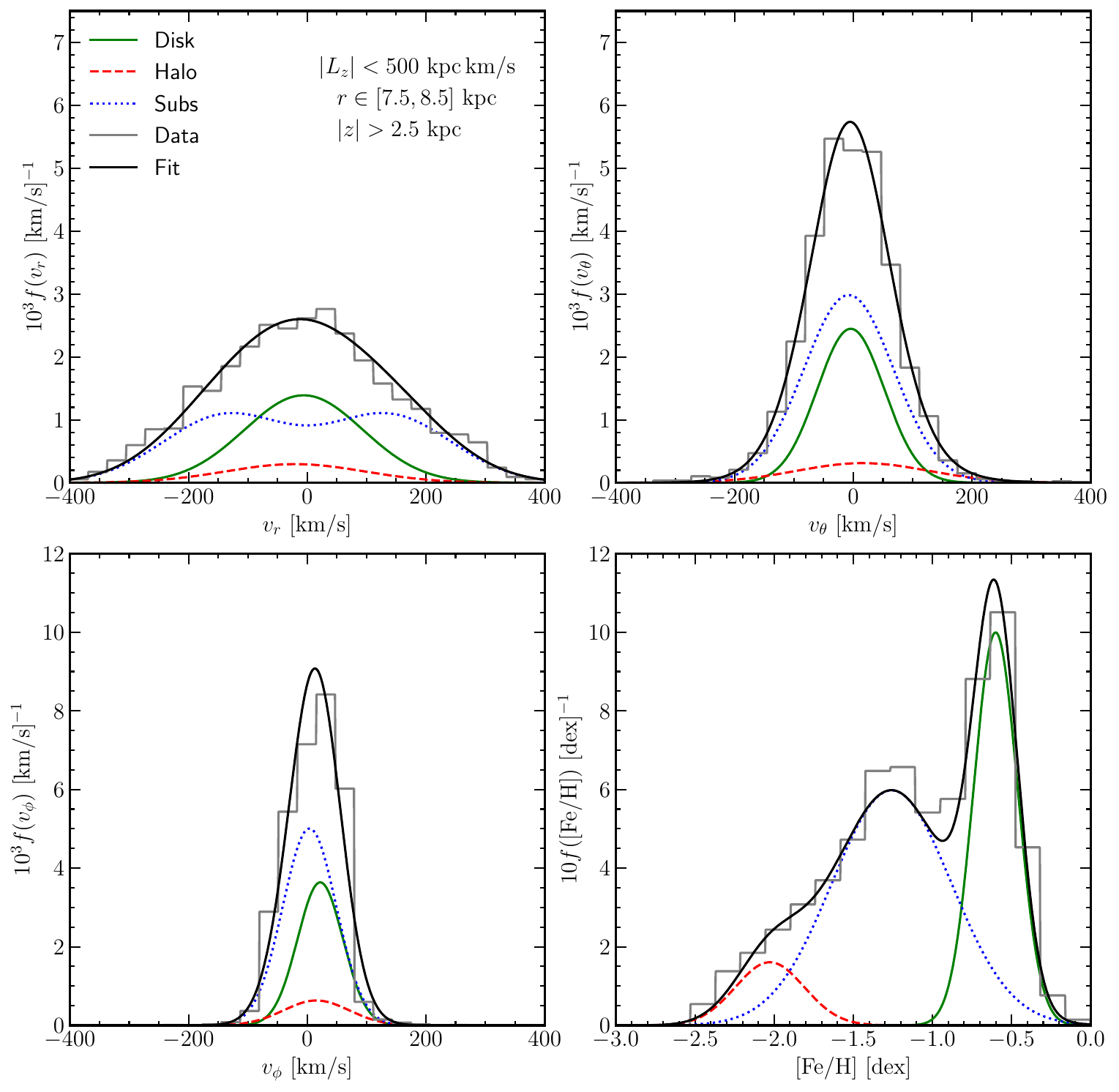}
\caption{The best-fit chemodynamical distributions of the sample in the region of $r\in[7.5,8.5]$~{\kpc} \& $|z|>2.5$~{\kpc} and $|L_{z}|<500$ \kpc\ \kms.}
\label{Lzcut}
\end{figure*}
\begin{equation}
\label{eq8}
\setlength{\abovedisplayskip}{2pt}
\setlength{\belowdisplayskip}{2pt}
\alpha_{\nu} = 0.25\nu^{2}-1.37\nu+1.69 \ ,
\end{equation}
\begin{equation}
\label{eq9}
\sigma_{\nu} = 0.2+\nu\times(\log_{10}M_{\rm peak}-\log_{10}M_{1}) \ .
\end{equation}
We set $\nu=-0.1$ in this benchmark model as in \citet{Necib_2019b}. 

\begin{figure*}[t]
\centering
\includegraphics[width=0.75\textwidth]{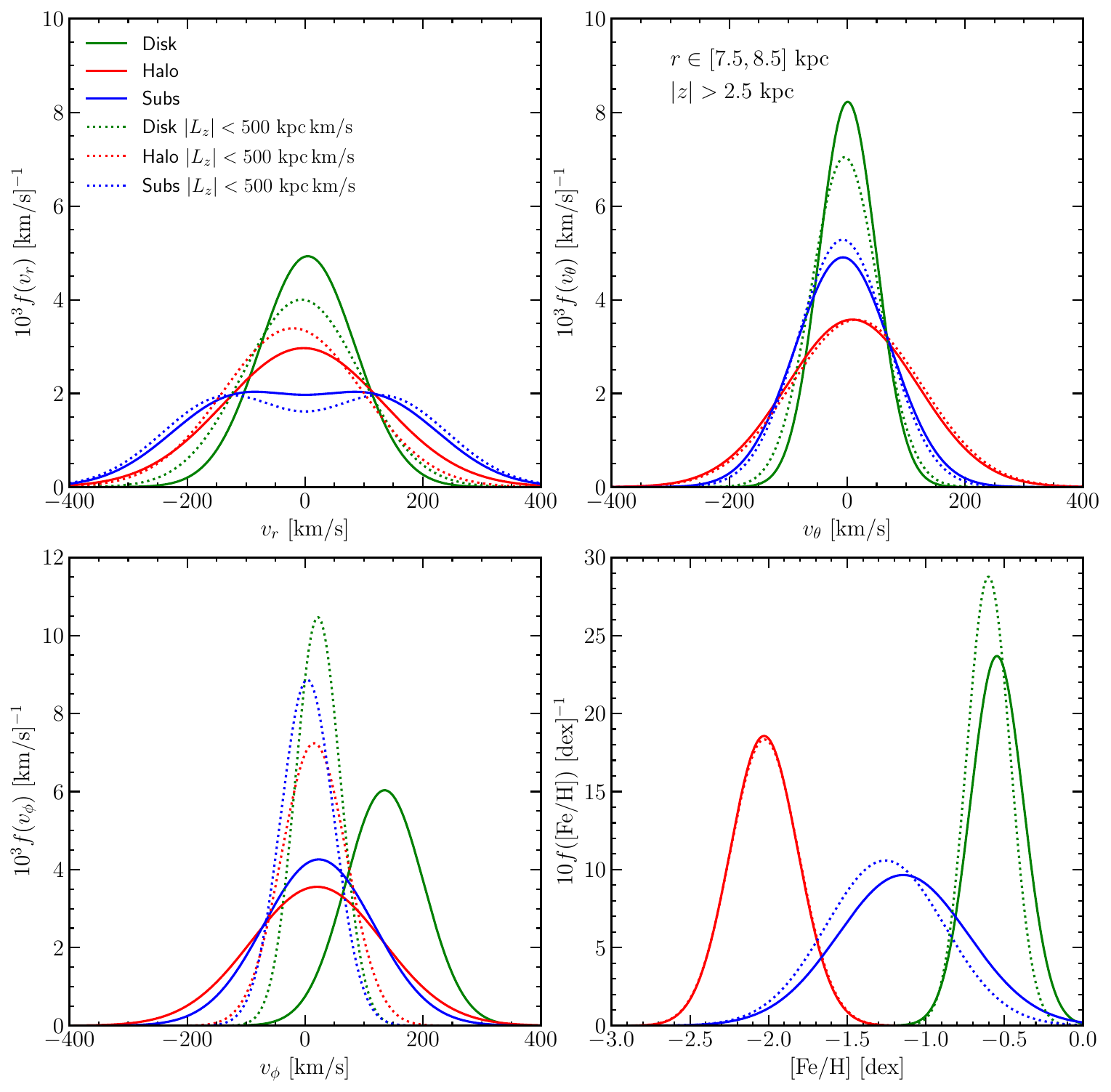}
\caption{The comparison of best-fit chemodynamical distributions of the sample in the region of $r\in[7.5,8.5]$~{\kpc} \& $|z|>2.5$~{\kpc} with (dotted lines) and without (solid lines) the $|L_{z}|<500$~\kpc~\kms\ cut. Unlike previous figures, the total area of each marginalized distribution (colored curves) is normalized to the same constant in each panel.}
\label{Lzcutcom}
\end{figure*}

Secondly, we utilize the \feh-$M_{\ast,\rm{total}}$ relation for dwarf galaxies in the MW investigated in \citet{Kirby_2013} to generate $\left< \feh\right>$:
\begin{equation}
\begin{aligned}
\left< \feh\right>&=(-1.69\pm0.04)\\
&+(0.30\pm0.02)\log_{10}\frac{M_{\ast,\rm{total}}}{10^{6}~M_{\odot}} \ .
\label{eq10}
\end{aligned}
\end{equation}
$\left< \feh\right>$ is randomly selected from the normal distribution with a median given by Eq.~\ref{eq10} and a dispersion of 0.17.

\begin{figure*}[t]
\centering
\includegraphics[width=\linewidth]{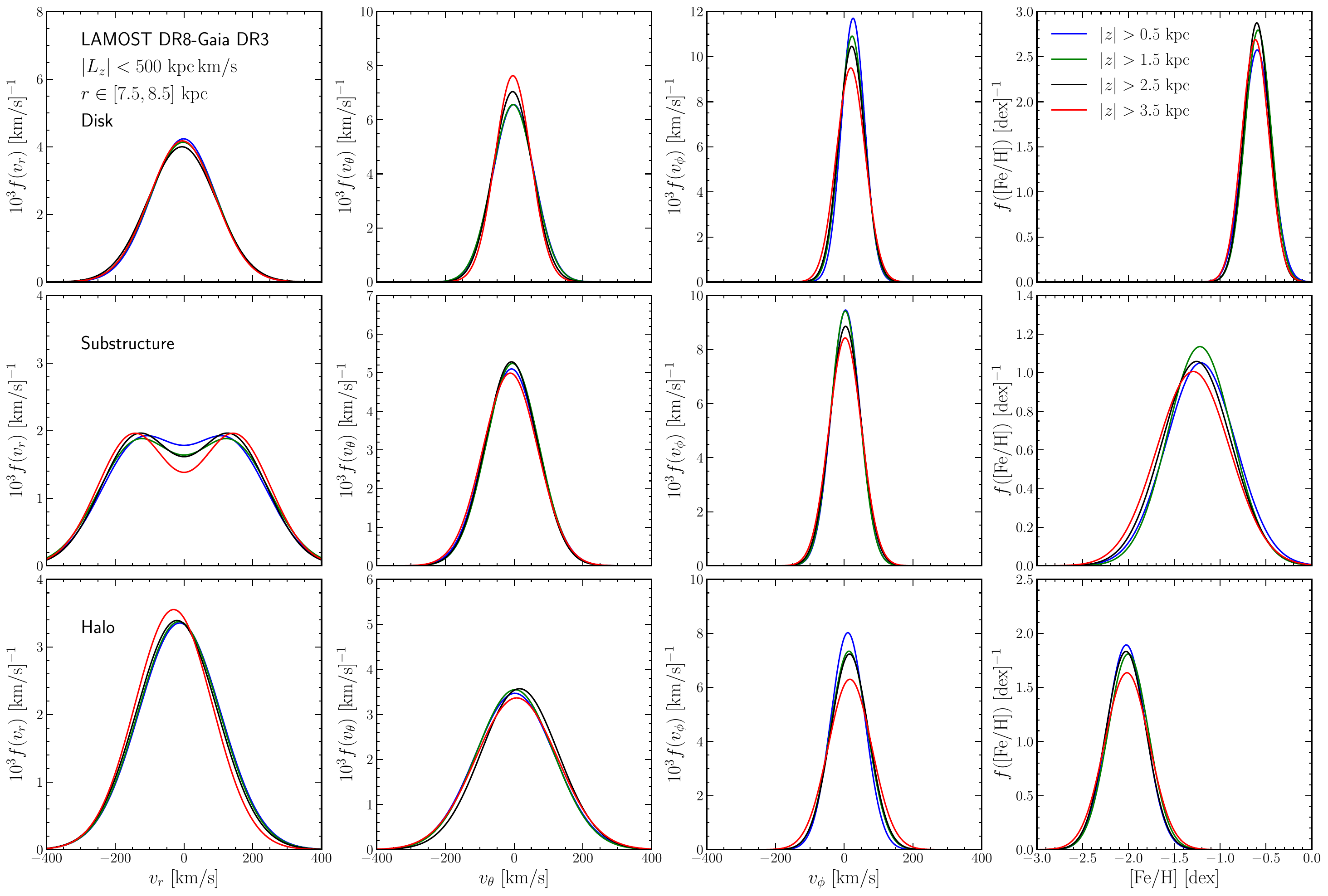}
\caption{Same as Fig.~\ref{vertical_variation_distribution} but for samples within $r\in[7.5,8.5]$~\kpc\ and an additional $|L_{z}|$ cut of $|L_{z}|<500$~\kpc~\kms. The distributions are roughly invariant even for the $v_{\phi}$ distribution of the disk component, as the angular momentum constraints excludes most disk stars and reduces its $v_{\phi}$ to a small value.}
\label{veritical_variation_distribution_Lz}
\end{figure*}

Finally, we repeat the process above to compute the uncertainties of $c_{s}\ \&\ c_{h}$ by selecting points around $\left<\feh\right>\sim-1.15$ and $\left<\feh\right>\sim-2.04$, representing the substructure and halo components, respectively. This benchmark model gives $\frac{c_{s}}{c_{h}}=0.05^{+0.10}_{-0.03}$ (16\%, 50\%, 84\% percentiles). Thus we can construct the Galactocentric velocity distribution of the local DM from Eq.~\ref{eq5} as follows:
\vspace{-2mm}
\begin{equation}
f_{\rm{dm}}(v)=\mathcal{Q}_{\rm{dm,s}}f_{\rm{s}}(v)+\mathcal{Q}_{\rm{dm,h}}f_{\rm{h}}(v) \ ,
\end{equation}
\vspace{-1mm}
where $\mathcal{Q}_{\rm{dm,s}}$ and $\mathcal{Q}_{\rm{dm,h}}$ are the mass fractions of DM in the substructure and isotropic halo components, respectively. In Eq.~\ref{eq5} the normalization requires

\begin{equation}
\label{eq12}
N(\mathcal{Q}_{\ast,h}+\frac{c_{s}}{c_{h}}\mathcal{Q}_{\ast,s}) = 1.
\end{equation}

From the above equation, $\mathcal{Q}_{\rm{dm,s}}=N\times\frac{c_{s}}{c_{h}}\mathcal{Q}_{\ast,s}=25^{+24}_{-15}\%$ and $\mathcal{Q}_{\rm{dm,h}}=1-\mathcal{Q}_{\rm{dm,s}}$, where $\mathcal{Q}_{\ast,s}=0.87$. The left panel of Fig.~\ref{helio_fv_mchi_sigmachi} shows the heliocentric speed ($v_{\rm{helio}}=|\boldsymbol{v}_{\rm{Gal}}-\boldsymbol{v}_{\odot,\rm{Gal}}|$) distribution of the substructure (blue dashed line) and halo (red dashed line) weighted by their mass fractions, and their combination (black solid line). For the comparison, the SHM is shown by the gray dashed line. \citet{Necib_2019} shows that the heliocentric speed distribution of dual model (consisting of the substructure and halo) shifts to the lower speed and has a sharper peak compared with the SHM, shown as the dark-blue shaded region in Fig.~\ref{helio_fv_mchi_sigmachi}. Our result agrees with them but shifts to an even lower speed by $\sim7\%$. The uncertainty of the heliocentric speed distribution, shown as the gray shaded region in Fig.~\ref{helio_fv_mchi_sigmachi}, is partially contributed by two parts: one is the uncertainty of the fractions of each component, the other is the uncertainty in the velocity distribution especially for the halo component. For the uncertainty due to the SMHM and {\feh}-$M_{\ast,\rm{total}}$ relations, we test to set the scatters in these two relations as zero. The resultant scatters in $\frac{c_{s}}{c_{h}}$ and $\mathcal{Q}_{\rm{dm,s}}$ are much smaller, which means that for uncertainties of the fractions ($\mathcal{Q}_{\rm{dm,s}}$), the uncertainty due to these two relations is dominant, though the median values are not changed.

\begin{figure*}[htbp]
\centering
\includegraphics[width=\linewidth]{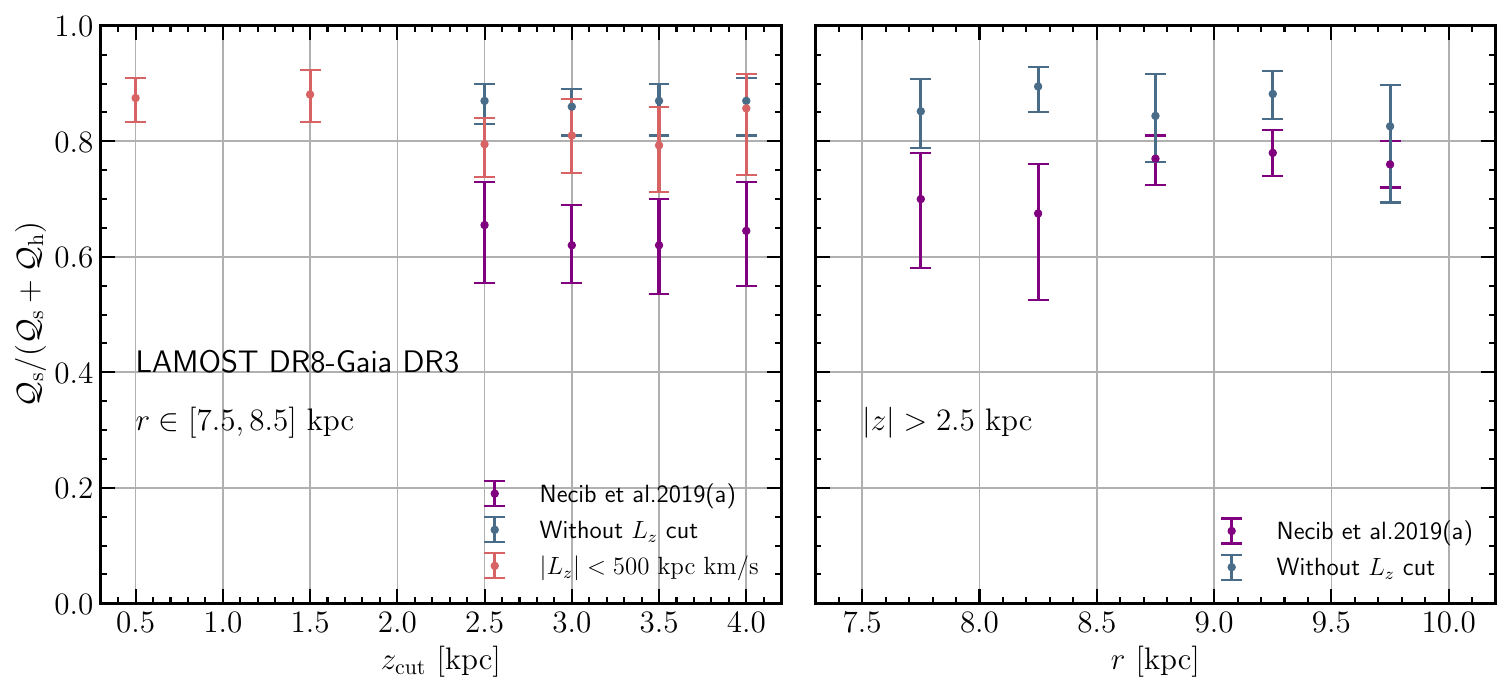}
\caption{Left: the vertical variation of the fraction of the substructure component relative to the nondisk components with $r\in[7.5,8.5]$~{\kpc}. Right: similar to the panel but for the radial variation.}
\label{spatial_variation}
\end{figure*}

\subsection{Experimental Detection Limits}
\label{Detection Limits}

Considering the scattering between the DM particle and the target nuclear with mass $m_{\chi}$ and $m_{T}$ respectively, the differential rate of scattering events is given by the following equation: 
\begin{equation}
\label{eq13}
\frac{\rm{d} \it R}{\rm{d} \it E_{R}}=\frac{\rho_{\chi}}{m_{\chi}m_{T}}\int_{v>v_{\rm{min}}}v\tilde{f}(\boldsymbol v)\frac{\rm{d}\sigma}{\rm{d}\it E_{R}}\rm{d}^{3}\it v \ ,
\end{equation}
where $E_{R}$ is the recoil energy deposited by the scattering, $\rho_{\chi}$ is the local DM density set as $0.3~\rm Gev/cm^{3}$ \citep{AMSLER20081,Baxter_2021} for both the SHM and the dual model, and $\tilde f(\boldsymbol{v})$ is the heliocentric velocity distribution\footnote{Critically, it should be the laboratory velocity distribution considering cyclically velocity of the earth with respect to the Sun. To simplify the computation, the time average of laboratory velocity over one period is utilized here, namely the heliocentric velocity.} of the local DM. $v_{\rm{min}}$ is the minimum velocity for a DM particle depositing the recoil energy $E_{R}$ defined as:
\vspace{-2.5mm}
\begin{equation}
\label{eq14}
v_{\rm{min}}=\sqrt{\frac{m_{T}E_{R}}{2\mu^{2}}}
\end{equation}
where $\mu$ is the reduced mass of the DM particle and target nuclear. 
The differential scattering cross section $\frac{\rm{d}\sigma}{\rm{d}\it E_{R}}$ is defined as: 
\vspace{-2.5mm}
\begin{equation}
\label{eq15}
\frac{\rm{d} \it \sigma}{\rm{d} \it E_{R}}=\frac{m_{T}A^{2}\sigma_{\rm{SI}}}{2\mu^{2}v^{2}}F^{2}(E_{R}) \ ,
\end{equation}
where $A$ is the atomic number of the target nuclear, $\sigma_{\rm_{SI}}$ is the spin-independent scattering cross section and $F(E_{R})$ is the nuclear form factor set as the Helm form \citep{Helm_1956}.
Then Eq.~\ref{eq13} can be written as:
\begin{equation}
\label{eq16}
\frac{\rm{d} \it R}{\rm{d} \it E_{R}}=\frac{\rho_{\chi}A^{2}\sigma_{\rm{SI}}}{2m_{\chi}\mu^{2}}F^{2}(E_{R})g(v_{\rm min}) \ ,
\end{equation}
where $g(v_{\rm min})$ is defined as: 
\begin{equation}
\label{eq17}
g(v_{\rm{min}})=\int_{v>v_{\rm{min}}}\frac{\tilde{f}(\boldsymbol v)}{v}\rm{d}^{3}\it v \ .
\end{equation}

Finally the total number of events is given by:
\begin{equation}
\label{eq18}
N_{e}=MT\int_{0}^{\infty}\phi(E_{R})\frac{\rm{d} \it R}{\rm{d} \it E_{R}}\rm{d}\it E_{R} \ ,
\end{equation}
where $M$ is the total mass of the target, $T$ is the exposure time and $\phi(E_{R})$ is the detector acceptance, i.e. the efficiency for different recoil energy, which differs for different experiments.

For a specific experiment, the number of events is the function of $m_{\chi}$, $\sigma_{\rm SI}$ and $\tilde{f}(\boldsymbol v)$. As an example, we utilize corresponding parameters of the Xenon direct detection experiment PandaX-4T \citep{PandaX4T_2021} with an exposure of 0.63 tonne $\times$ year. Applying the SHM, we convert the published exclusion curve of PandaX-4T into $N_{e}$ vs $m_{\chi}$ with the efficiency curve given in \citet{PandaX4T_2021}. Then applying the DM velocity distribution deduced in Section.~\ref{DM Distribution}, we re-compute the cross section limit resulting in the same $N_{e}$ vs $m_{\chi}$. All points have $m_{\chi}\gtrsim5$~Gev due to the low signal efficiency for low mass DM particles.

The right panel of Fig.~\ref{helio_fv_mchi_sigmachi} shows the 90\% C.L. direct detection limits of the DM substructure (blue dashed), the DM halo (red dashed) and their total contribution (black solid). When computing the $g(v_{\rm{min}})$ we use the fraction weighted velocity distribution of the substructure and halo components. Compared with the SHM (gray dashed), the dual model results in a weaker limit for lower mass DM due to the softer DM kinetic energy distribution, and a slightly stronger limit for DM mass above 150 GeV for which the signal acceptance increases for a given set of selection cuts.

\section{Discussion}
\label{Sec6: Discussion}

\begin{figure*}[t]
\centering
\includegraphics[width=\textwidth]{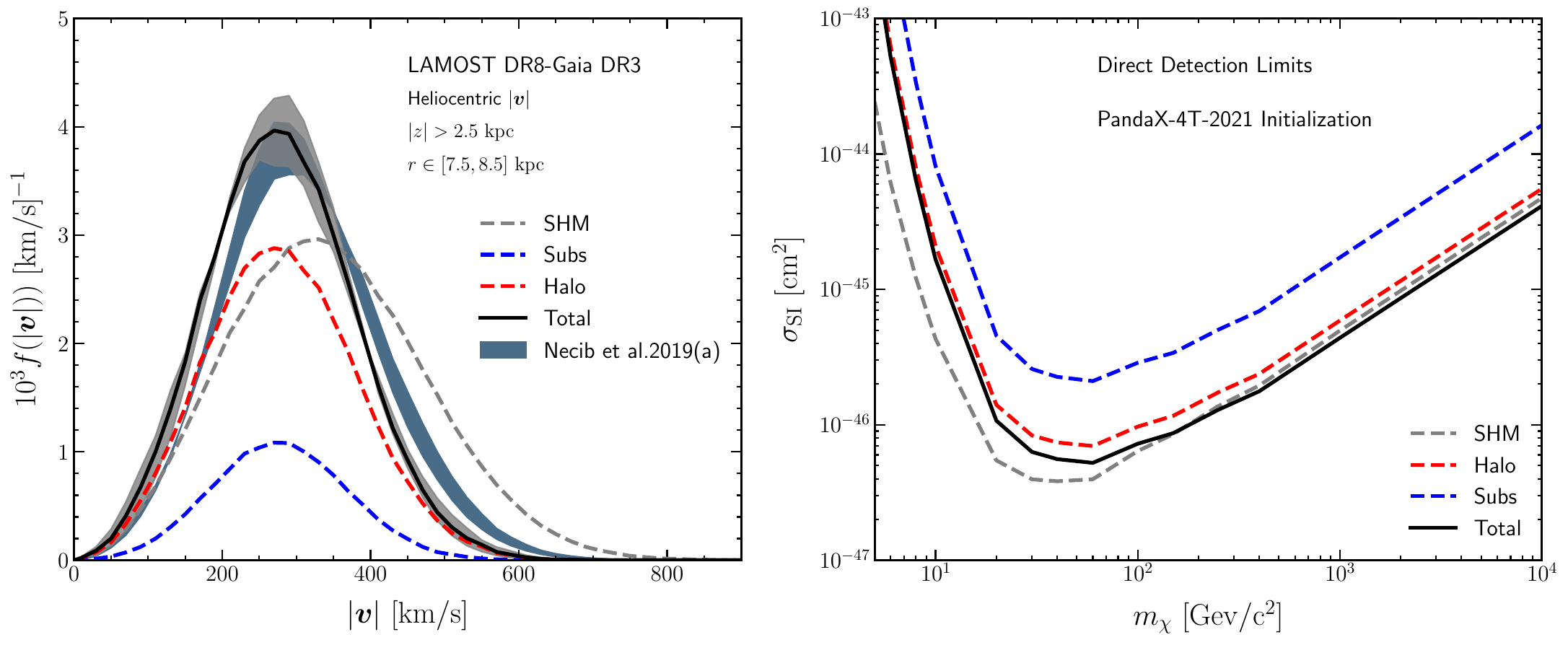}
\caption{Left: the heliocentric speed distributions of the best-fit substructure (blue dashed line) and halo (red dashed line) components, weighted by their relative fractions in the Solar neighbourhood. The combination of the two is shown as the black solid line, while the gray shaded region displays the uncertainty. The gray dashed line represents the SHM. The dark-blue shaded region shows the result from \citet{Necib_2019}. Right: 90\% C.L. upper DM direct detection limits on the cross section ($\sigma_{\rm SI}$) vs. DM mass ($m_{\chi}$) plane, calculated based on the DM direct detection experiment PandaX-4T \citep{PandaX4T_2021}. The representation of each line is the same as the left panel.}
\label{helio_fv_mchi_sigmachi}
\end{figure*}

\subsection{Methods of Decreasing the Contribution of Disk Stars}
\label{Sec6.1: Methods of Decreasing the Contribution of Disk Stars}
In this work our main interest is the nondisk stars, i.e. the substructure and main halo components. The disk-dominance could lead to uncertainty for other two components. There are many methods to remove the disk stars, such as constraints on the spatial volume, the chemical abundance, the kinematic information etc. The vertical height cut is applied in Section~\ref{Sec3: Stellar Distribution in the Chemodynamical Space}, and an additional angular momentum ($L_{z}$) constraint is explored in Section~\ref{Sec4: Angular Momentum Constrain}. The former shows the invariance of the original chemodynamical distribution of the nondisk components in $z$ direction. The latter can efficiently reduce the disc dominance while keeping the applicability of the GMM, though it changes the $v_{\phi}$ distribution of the nondisk components. The invariant results of these two methods demonstrate the reliability of extrapolating our results to the Solar vicinity. Here we test other two methods, i.e. the metallicity cut and Toomre diagram cut, to decrease the contribution of disk stars and assess their influence.

\citet{Wu_2022} applied an effective selection criteria ($|z|>5$~{\kpc} \& $ {\feh}<-0.5$) to remove possible disk stars. As our interested region is the Solar neighbourhood with lower vertical height, we attempt to cut the stellar sample with a more strict criteria of \feh\ $\lesssim-1$ to exclude most of possible disk stars. We then fit the sample using a two-component model with only the substructure and halo components. However, as the mean \feh\ of the substructure component is $\sim - 1.2$~dex, this metallicity cut would also exclude nearly half of the substructure stars. The GMM method failed to fit the stellar sample, as the peak metallicity of the substructure component would prefer a value beyond the upper metallicity boundary of $- 1$~dex.

\begin{figure}[t]
\centering
\includegraphics[width=\linewidth]{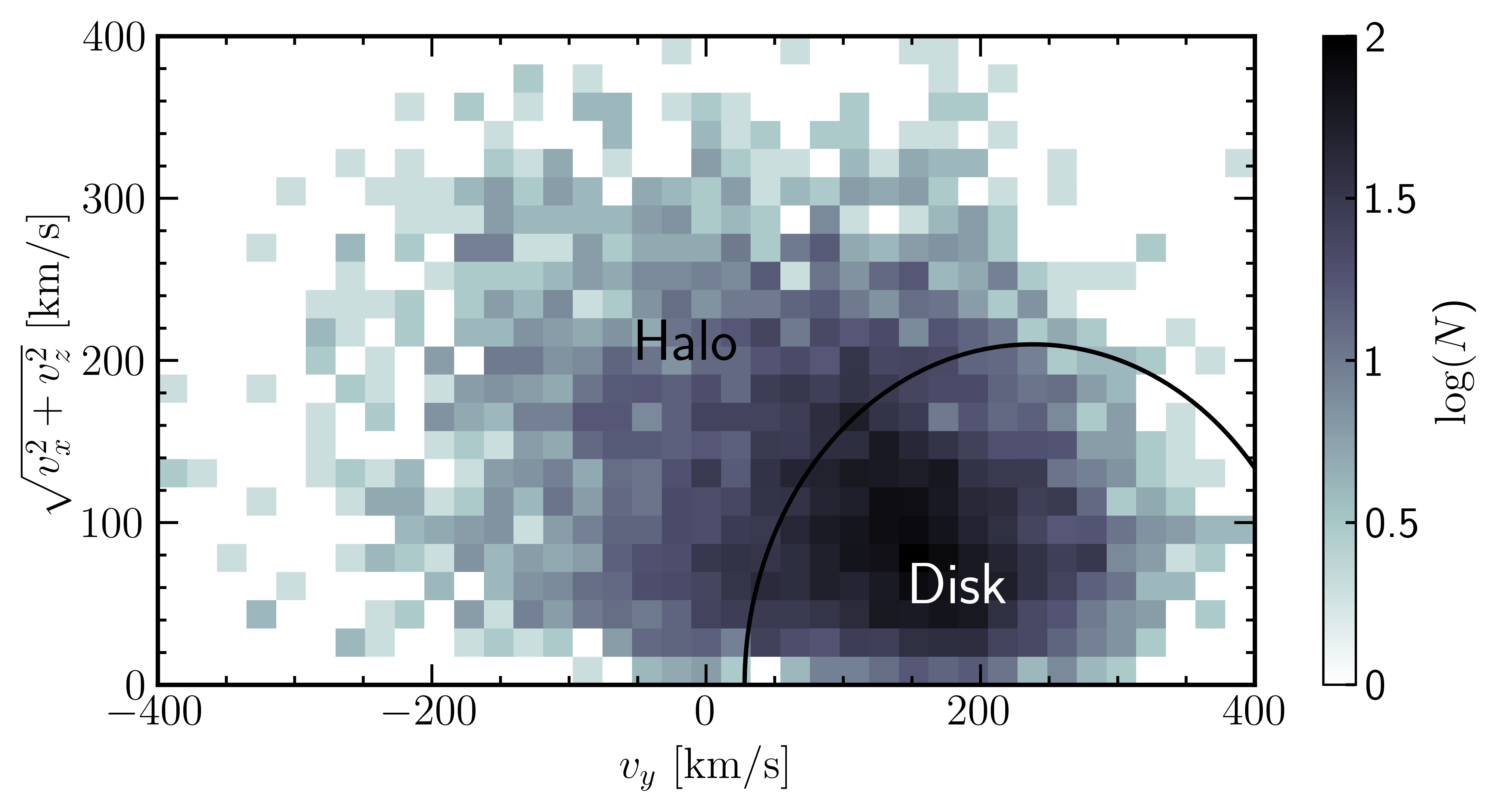}
\caption{Toomre diagram for the sample in the region of $r\in[7.5,8.5]$~\kpc\ $\&$ $|z|>2.5$~{\kpc}. The black circle represents $|\boldsymbol{v}-\boldsymbol{v}_{lsr}|=210$~\kms\ where $\boldsymbol{v}_{lsr}=(0,238,0)$~\kms, which presumably separates halo stars from disk stars.}
\label{toomre_diagram}
\end{figure}

In Section~\ref{Sec4: Angular Momentum Constrain} we show one of the possible kinematic constraints to decrease the contribution of disk, which supports our results in Section~\ref{Sec3: Stellar Distribution in the Chemodynamical Space} and also extrapolates our results to the Galactic plane. Another commonly applied kinematic cut to select the GES stars is the Toomre diagram as shown in Fig.~\ref{toomre_diagram}. After excluding disk stars from the Toomre diagram, the chemodynamical distribution of our sample shows a nonnegligible contribution of metal-rich stars with peak \feh\ $\sim-0.6$~dex. Besides, $v_{r}\ \&\ v_{\phi}$ distributions of the remained sample show prominent asymmetry as the cut is asymmetric in velocity space, which excludes more stars with prograde rotation. Thus, the GMM again failed to fit the sample after the Toomre diagram cut due to such an asymmetry. Though we can separate the bimodal Gaussian distribution into two Gaussian distributions with different fractions, the fitting only gives consistent results with Section~\ref{Sec3: Stellar Distribution in the Chemodynamical Space} in the means of velocities. It may bias the resultant velocity distributions of the nondisk components.

The asymmetry of $v_{r}\ \&\ v_{\phi}$ distribution indicates that the meal-rich peak might correspond to the hot thick disk and the Nyx stream \citep{Necib_2020}. Furthermore, \citet{Donlon_2023} combined this kinematic cut with an additional chemical abundance cut~\citep{Das_2020,Buder_2022,Belokurov_2022al} and obtained a local stellar halo sample. Their selection requires more chemical abundance information such as $[\rm{Al/Fe}],\ [\rm{Mg/Mn}],\ [\rm{C/Fe}]$ besides \feh. In the future, based on more chemical abundance information from LAMOST, we might obtain more detailed structures of the local stellar halo. 

\subsection{Degeneracy between Substructure and Halo}
\label{Sec6.2: Degeneracy between Substructure and Halo}

In Section~\ref{Parameter Degeneracy} we investigate the potential degeneracy between the substructure and halo components especially for the radial velocity and metallicity distribution. As the halo component contributes a small fraction to the total sample, we also tried a two-component model including only the disk and substructure components to fit the sample, and compare with the results from the three-component model. 
 
Besides in the regions of $r\in[8.5,9.0]$~\kpc\ and $r\in[9.5,10.0]$~\kpc\ when $|z|>2.5$~\kpc, the halo component has a mean metallicity $\mu_{\feh}\sim-1.5$~dex. The different $\mu_{\feh}$ from that in other radial bins could be caused by the strong degeneracy with the substructure component. Therefore we firstly apply the two-component model in these two regions and compare with the three-component model. 

The Bayesian information criterion \citep[BIC, ][]{Schwarz_1978} is utilized to quantify which model performs better. The BIC is defined as:

 \begin{equation}
 \label{eq19}
 \rm{BIC}=\it k\ln N-\rm2\times\ln\mathcal{L}_{MAP} \ ,
 \end{equation}
where $k$ is the number of free parameters in the model, $N$ is the size of the sample and $\mathcal{L}_{\rm MAP}$ is the maximum of the likelihood function deciding the best-fit parameters. Fitting to the same sample, these two models share the same $N$ while the two-component model contains 23 free parameters in total.

We calculate the BIC for each model and find $|\rm{BIC_{2com}}-\rm{BIC_{3com}}|\lesssim10$, which means both models fit the sample statistically similarly. Nevertheless, for other radial bins, $\rm{BIC_{2com}}-\rm{BIC_{3com}} \gtrsim 100$, which means that the three-component model performs better and the contribution from the halo component is not trivial.

One of possible contributions to the degeneracy is from other substructures with similar chemical abundance and kinematic properties as the substructure and halo components. As mentioned in \citet{Wu_2022} the most significant contaminant is the Sgr stream with $\mu_{\feh}\sim-1.3$ dex \citep{Yang_2019}, which has similar metallicity as the substructure component. Nevertheless, as we focus on the Solar neighbourhood with $r < 10$ kpc, the contribution of the Sgr stream is believed to be negligible \citep{Yang_2019}. Their results with the Sgr stream removed showed that the radial motion of the GES component, i.e. the substructure component in our work, is affected by other substructures. With most substructures removed, the radial velocity distribution of the GES component shows stronger bimodal distribution.  Note that, the hot thick disk and the Nyx stream discussed in \citet{Necib_2020} might contribute to the disk and the substructure components, which need more careful studies in the future.

We try to fit the data with an additional halo substructure i.e. a four-component model. Comparing the BIC of three-component model and four-component model, we find $\rm{BIC_{4com}-BIC_{3com}\sim-79}$, which means considering another substructure may fit the data slightly better. However we found the fractions of the disk and isotropic halo components are 0.5 and 0.06, respectively, which are similar to those ($\mathcal{Q}_{d}= 0.55$, $\mathcal{Q}_{h}= 0.06$) of the three-component model. The only change seems to be that the old bisymmetric substructure component (with two overlapped Gaussians in $v_r$) is decomposed into an isotropic halo with $\mu_{\feh}\approx -1.0$ and a new bisymmetric substructure with larger $\mu_{r}\sim200$ {\kms}, smaller $\sigma_{r}\sim88$ {\kms} and almost no overlap between the two Gaussians in $v_r$. This is kind of expected due to the similar Gaussian feature of these two models. It seems that we cannot identify other halo substructures such as the Nyx stream from the current data, which could be partially due to the large fraction of the disk stars ($\sim$50\%). A pure halo sample would be better to find other halo substructures.

As a summary, we list the possible major causes of the degeneracy between the substructure and halo components as following:
\begin{itemize}
\item {The model setup. The bisymmetric Gaussian substructure is possible to be decomposed into a new isotropic halo with the similar mean metallicity and a small fractional substructure with larger $\mu_{r}$ and smaller $\sigma_r$. The new isotropic halo may then be degenerate with the halo and disk components.}

\item {The large contribution of the disk component. The disk fraction in our sample is about 50\%, while the obtained halo component is only 6\%. The small fraction of the halo component makes it difficult to be identified. A purer halo sample would be better to identify the nondisk components and alleviate the degeneracy.}

\item {The metalicity floor (-2.5 dex) of the sample. In our results, the halo component has $\mu_{\feh}= -2.04$ and $\sigma_{\feh}= 0.21$. The floor may slightly influence the identification of the halo component, though the fraction of giants with {\feh}$<$ -2.5 is estimated to be small.}

\item {Other accreted substructures not identified in the data and model. Many accreted substructures such as the Nyx stream, have been identified in the stellar halo. Although it is difficult to identify such substructures in the current data, it may lead to some degeneracy if their contribution is nontrivial.}
\end{itemize}

These degeneracies could be reduced in the future utilizing a purer halo sample and combining more kinematic (e.g. $L_z$, energy) and chemical (e.g. $\alpha-$abundance, [Al/Fe]) information.

\subsection{Limitation of Tracing DM Velocity Distribution with Stars}
The correlation in velocity distribution between the accreted stars and DM is the basic assumption in this work, which is still being debated. Such a correlation is found for those accreted from luminous satellite galaxies. However, \citet{Wang_2011} and \citet{Necib_2019} showed that a nonnegligible fraction of DM of Milky Way-like galaxies could be accreted from smooth accretion of dark satellites, which cannot be traced by accreted stars. This limitation is also mentioned in \citet{Bozorgnia_2019} to explain the weak correlation in the velocity distributions of metal-poor stars and DM particles in \texttt{Auriga} simulations. We tried to investigate such correlation in TNG-50 simulations \citep{Pillepich_2019,Nelson_2019} and found a rough correlation between stars and DM accreted from the same satellite. However, the resolution of TNG-50 simulations is still not enough for most Milky Way-like galaxies to check the velocity distribution for accreted particles in the Solar neighbourhood. Higher resolution for Milky Way-like galaxies is needed to further investigate the correlation between the stars and DM. This correlation may also depend on the merger orbit of the satellite galaxy.

The benchmark model of converting the relative contribution of the substructure in the stellar sample to that in DM is very sensitive to the metallicity of each stellar component. Different spectroscopic surveys have different selection effects thus different metallicity distribution of the stellar sample. It may contribute to the difference between our results and \citet{Necib_2019}. Besides, this conversion also depends on the SMHM relation of the satellites which varies for different simulations. All these could result in additional uncertainties in the relative contribution from the DM substructure.

Besides the contribution of the radial merger event of GES happened around 8$-$10 Gyr ago \citep{Sahlholdt_2019,Bignone_2019,Bonaca_2020}, other recent merger events may also have nonnegligible influence on the local DM density and velocity distribution. \citet{OHareCiaranAJ_2020} computed the impact of the S1 stream on WIMP and axion detectors, which passes through the Solar neighbourhood on a low inclination and counter-rotating orbit, with a progenitor mass comparable to the present-day Fornax dwarf spheroidal. \citet{Donlon_2022} identified three radial merger events, i.e. Virgo radial merger, Nereus and Cronus, using local halo dwarf stars. They thought stars comprising the GES velocity structure are a combination of those components. \citet{smithorlik_2023} studied the effect of the Large Magellanic Cloud (LMC) on the DM distribution in the Solar neighbourhood utilizing simulations of Milky Way analogues. They found that DM particles in the Solar neighbourhood originating from the LMC analogue dominate the high velocity tail of the local DM velocity distribution, and the native DM particles of the MW in the Solar vicinity are boosted to higher speeds as a result of a response to the LMC’s motion.

\section{Conclusion}
\label{Sec7: Conclusion}


In this work, we attempt to study the local DM velocity distribution based on the correlations in velocity distribution between stars and DM belonging to different components. We apply the GMM to separate the disk, the halo substructure and the main halo components in the chemodynamical space utilizing K giants of LAMOST DR8 cross-matched with {\gaia} DR3. The substructure component is radially anisotropic possibly related to the GES debris flow, while the halo component is referred to as the isotropic halo accreted earliest and thus virialized. Our main results are derived from the subsample within $r\in[7.5,10]$~\kpc\ \&\ $|z|>2.5$~\kpc. We also explore the spatial variation of the best-fit distributions. We try to apply an additional angular momentum constraint to reduce the dominance of the disk component. The main results are consistent and are invariant with the vertical cut, which verifies the robustness of extrapolating our main results to the Solar vicinity. Based on the stellar chemodynamical distributions, we compute the velocity distribution of the dual DM, i.e. the substructure and halo components, applying the \feh-$M_{\ast,\rm{total}}$ relation and the SMHM relation. This modified DM velocity distribution is different from the SHM, resulting in different exclusion curves ($\sigma_{\rm SI}$ vs. $m_{\chi}$) for DM direct detection experiments.


Our main results are as follows:
\begin{itemize}
\item {The GMM performs well in separating stars belonging to different components in the chemodynamical space. For our K giant stars within $r\in[7.5,8.5]$~\kpc\ $\&$ $|z|>2.5$~{\kpc}, the substructure component, i.e. the GES debris flow, contributes about 87\% of the nondisk stars ($\mathcal{Q}_{s}/(\mathcal{Q}_{h}+\mathcal{Q}_{s})=0.87^{+0.03}_{-0.04}$). This dominant stellar halo component is highly radially anisotropic with an anisotropy parameter $\beta=1-\frac{\sigma_{\theta}^{2}+\sigma_{\phi}^{2}}{2\sigma_{r}^{2}} = 0.71$, which is smaller than some previous works \citep[e.g.][]{Belokurov_2018b,Myeong_2018,Bird_2019,Wu_2022}. The halo component referred to as the virialized halo is more isotropic with $\beta=0.23$. Both components show a slightly prograde rotation with $\mu_{\phi}\sim20$~\kms. The median metallicities for the substructure and halo components in our sample are $-1.15$ and $-2.04$, respectively. The substructure component is slightly more metal-rich and the halo component is more metal-poor than the corresponding compnents in \citet{Necib_2019} and \citet{Wu_2022}.}

\item {The best-fit chemodynamical distributions of the nondisk components are almost invariant for subsamples with different vertical cuts or different radial ranges in $r\in[7.5,10]$~\kpc\ and $|z|>2.5$~\kpc. Furthermore we extrapolate our results to the Galactic plane with an additional angular momentum ($L_{z}$) constraint to reduce the dominance of the disk component. The main results with the angular momentum constraint, e.g. $\mathcal{Q}_{s}/(\mathcal{Q}_{h}+\mathcal{Q}_{s})$, are consistent with those from samples without the $L_{z}$ cut. These tests demonstrates the robustness of extrapolating our main results to the Solar vicinity. 
}

\item {Some parameters in the GMM are degenerate, especially between the parameters of metallicity ($\mu_{\feh}$, $\sigma_{\feh}$) and the fractions ($\mathcal{Q}_{s}$, $\mathcal{Q}_{h}$) of the substructure and halo components. These correlations imply the degeneracy between the substructure and halo components, which is partly due to the relatively small contribution of the halo component. Nevertheless, the BIC test indicates that in most regions (except r$\in[8.5,9.0]$~\kpc\ and $[9.5,10]$~\kpc), the three-component model is preferred over a two-component model, which shows that the halo component is nonnegligible.
}

\item {Utilizing the \feh-$M_{\ast,\rm{total}}$ relation and the SMHM relation, we convert the relative contribution of the substructure component to the total halo stars ($\mathcal{Q}_{s}/(\mathcal{Q}_{h}+\mathcal{Q}_{s}) \sim 85\%$) to the relative contribution in DM of $25_{-15}^{+24}\%$ in the Solar neighbourhood. As a comparison, \citet{Necib_2019} indicated the substructure contributes 60\%-80\% of the local total halo stars and $42^{+26}_{-22}\%$ of DM.
}

\item {Considering the contributions from both the substructure and halo DM, the local DM heliocentric velocity distribution shifts to the lower speed and has a sharper peak compared to the SHM. For the Xenon direct detection experiment PandaX-4T, the dual DM model indicates a significant larger cross section $\sigma_{\rm{SI}}$ for low mass DM of $m_{\chi}\lesssim150$~Gev.
}
\end{itemize}

\section*{Acknowledgement}
\label{Acknowledgement}
The research presented here is partially supported by the National Key R\&D Program of China under grant No. 2018YFA0404501; by the National Natural Science Foundation of China under grant Nos. 12025302, 11773052, 11761131016, 11333003, 12103031; and by the ``111'' Center of the Ministry of Education under grant No. B20019. This project was developed in part at the LAMOST-Gaia Sprint 2022, supported by the National Natural Science Foundation of China (NSFC) under grants 11873034 and U2031202. R.G. is supported by Initiative Postdocs Supporting Program (No. BX2021183), funded by China Postdoctoral Science Foundation. J.S. acknowledges support from a Newton Advanced Fellowship awarded by the Royal Society and the Newton Fund. J.L. is supported by NSFC (No. 1209060), the Hongwen Foundation in Hong Kong, and Tencent and New Cornerstone Science Foundation in China. X.-X.X. and L.Z. acknowledge the support from National Key Research and Development Program of China No. 2019YFA0405504, National Natural Science Foundation of China (NSFC) under grants No. 11988101, CAS Project for Young Scientists in Basic Research grant No. YSBR-062 and YSBR-092, and China Manned Space Project with No. CMS-CSST-2021-B03.

Guoshoujing Telescope (the Large Sky Area Multi-Object Fiber Spectro-scopic Telescope, LAMOST) is a National Major Scientific Project built by the Chinese Academy of Sciences. Funding for the project has been provided by the National Development and Reform Commission. LAMOST is operated and managed by the National Astronomical Observatories of China, Chinese Academy of Sciences.

This work has made use of data from the European Space Agency (ESA) mission {\gaia} (\url{https://www.cosmos.esa.int/gaia}), processed by the {\gaia} Data Processing and Analysis Consortium (DPAC,\url{https://www.cosmos.esa.int/web/gaia/dpac/consortium}). Funding for the DPAC has been provided by national institutions, in particular the institutions participating in the {\gaia} Multilateral Agreement.

This work made use of the Gravity Supercomputer at the Department of Astronomy, Shanghai Jiao Tong University.

\appendix
\label{Appendix}

\begin{figure*}[htbp]
\centering
\includegraphics[width=\textwidth]{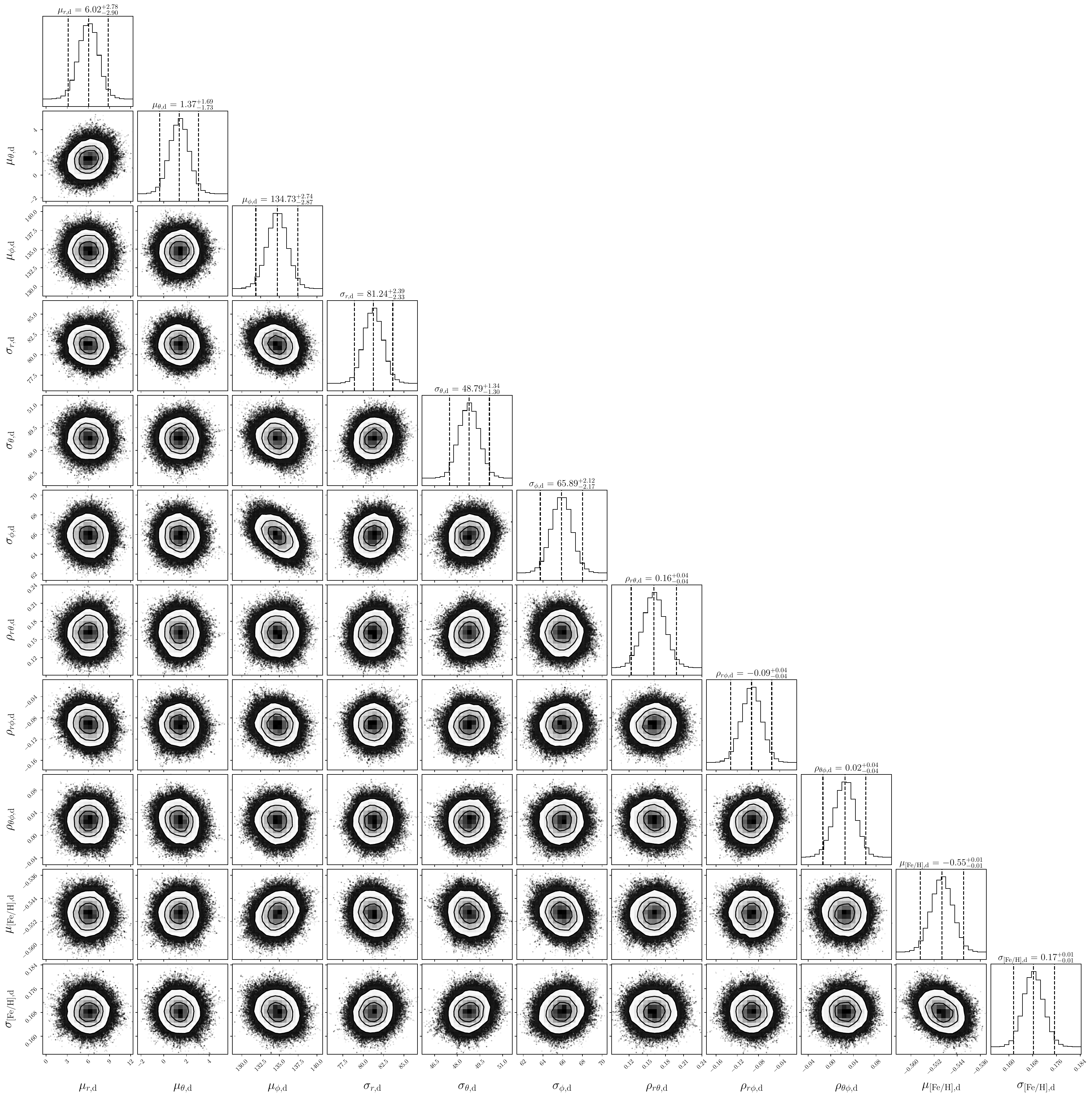}
\caption{The corner plot of the model parameters of the disk component for the sample in the region of $r\in[7.5,8.5]$~{\kpc} \& $|z|>2.5$~{\kpc}.}
\label{corner_disk}
\end{figure*}

\begin{figure*}[htbp]
\centering
\includegraphics[width=\textwidth]{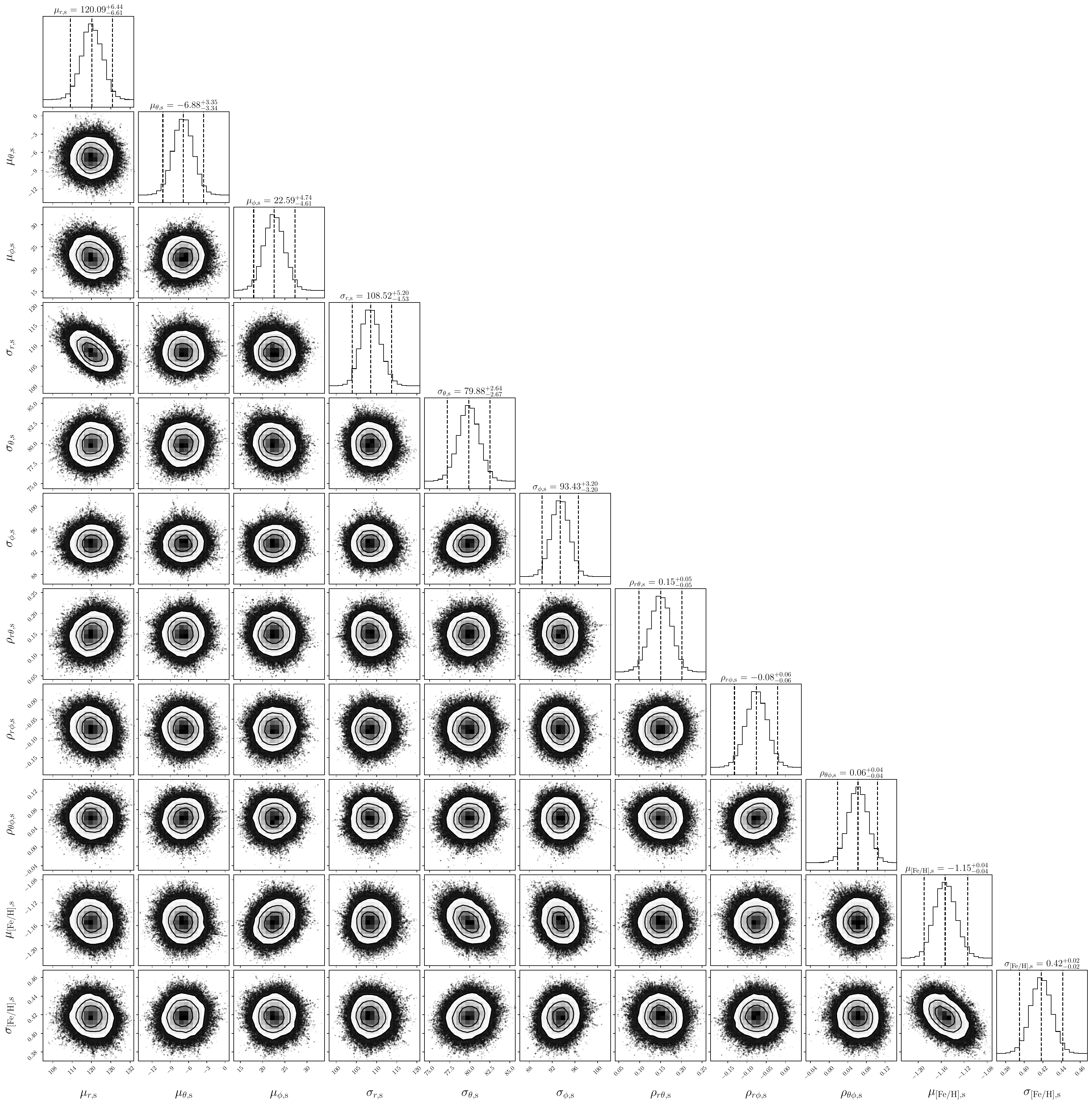}
\caption{The corner plot of the model parameters of the substructure component for the sample in the region of $r\in[7.5,8.5]$~{\kpc} \& $|z|>2.5$~{\kpc}.}
\label{corner_subs}
\end{figure*}

\begin{figure*}[htbp]
\centering
\includegraphics[width=\textwidth]{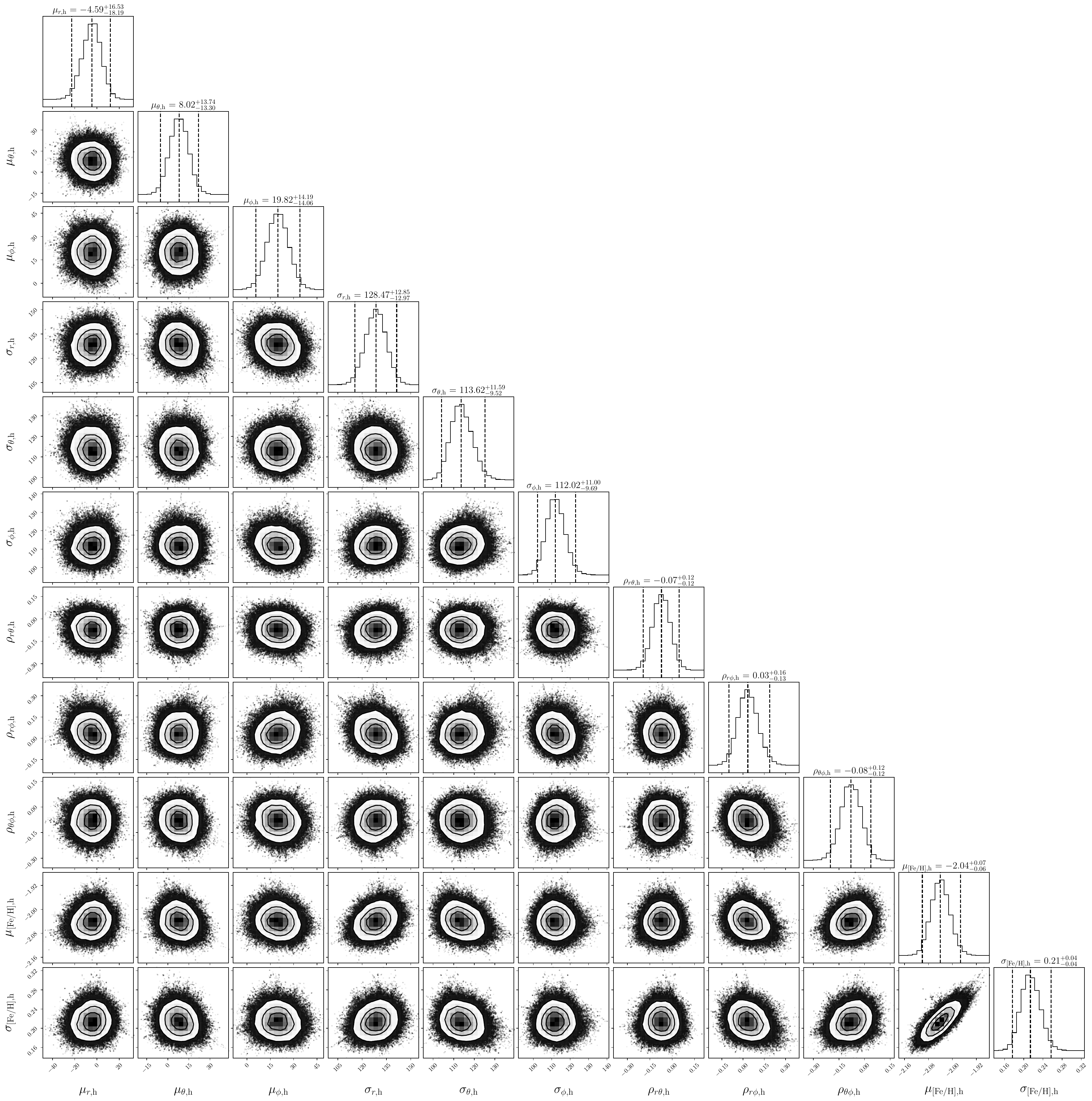}
\caption{The corner plot of the model parameters of the halo component for the sample in the region of $r\in[7.5,8.5]$~{\kpc} \& $|z|>2.5$~{\kpc}.}
\label{corner_halo}
\end{figure*}

We provide more details of our sample and the GMM results in this appendix. Table~\ref{table2} provides the characteristic uncertainties of our sample and the comparison with previous studies. In Tables~\ref{table3}-\ref{table5} we provide posterior distribution of parameters of the GMM for the disk, substructure and halo component, respectively. The corner plots for the sample in the region $r\in[7.5,8.5]$~{\kpc} \& $|z|>2.5$~{\kpc} are displayed in Figs.~\ref{corner_disk}-\ref{corner_halo}. 

\begin{deluxetable*}{ccccccccc}[h]
\centering
\tablecaption{Comparison of characteristic uncertainties of the previous samples and ours.}
\label{table2}
\tablewidth{0pt}
\setlength{\tabcolsep}{1mm}{
\tablehead{
    \colhead{Sample}&\colhead{$\delta T_{eff}$}&\colhead{$\delta\log g$}&\colhead{$\delta\rm [Fe/H]$}&\colhead{$\delta v_{los}$}&\colhead{$\delta\varpi/\varpi$}&\colhead{$\delta d_{\rm carlin}/d_{\rm carlin}$}&\colhead{$\delta d_{\rm xue}/d_{\rm xue}$}&\colhead{$\delta d_{\rm phot}/d_{\rm phot}$}\\
\colhead{}&\colhead{K}&\colhead{dex}&\colhead{dex}&\colhead{km/s}&\colhead{}&\colhead{}&\colhead{}&\colhead{}}
\startdata
LAMOST K giants (this paper) &$61.6$&  $0.101$ & $0.059$ & $4.4$ & $21.62\%$ & $18.19\%$ & $23.29\%$ &$-$\\
Selected K giants (this paper)&$68.52$&  $0.112$ & $0.066$ & $4.52$ & $24.78\%$ & $18.31\%$ & $20.38\%$ &$-$\\
MS stars N19a&$-$&  $-$ & $-$ & $<50$ &$-$& $-$ & $-$ & $10\sim20\%$ \\
SEGUE K W22&$-$&  $-$ & $0.12$ & $2$ & $-$ & $-$ & $16\%$ &$-$\\
LAMOST K W22&$-$&  $-$ & $0.14$ & $7$ & $-$ & $-$ & $13\%$ &$-$\\
SDSS BHB W22&$-$&  $-$ & $0.21$ & $4$ & $-$ & $-$ & $5\%$ &$-$\\
\enddata}
\end{deluxetable*}

\begin{deluxetable*}{cccccccccccc}[h]
\label{table3}
\centering
\tablecaption{Parameters for the GMM of the disk component with and without $L_{z}$ cut in the region of $r\in[7.5,8.5]$ \kpc\ \&\ $|z|>z_{cut}$.}
\tablewidth{0pt}
\setlength{\tabcolsep}{1mm}{
\tablehead{
	\colhead{$z_{cut}$}&\colhead{$\mu_{r}$}&\colhead{$\mu_{\theta}$}&\colhead{$\mu_{\phi}$}&\colhead{$\sigma_{r}$}&\colhead{$\sigma_{\theta}$}&\colhead{$\sigma_{\phi}$}&\colhead{$\rho_{r\theta}$}&\colhead{$\rho_{r\phi}$}&\colhead{$\rho_{\theta\phi}$}&\colhead{$\mu_{\feh}$}&\colhead{$\sigma_{\feh}$}\\
\colhead{kpc}&\colhead{\kms}&\colhead{\kms}&\colhead{\kms}&\colhead{\kms}&\colhead{\kms}&\colhead{\kms}&\colhead{}&\colhead{}&\colhead{}&\colhead{dex}&\colhead{dex}}
\decimals
\startdata
\multicolumn{12}{c}{With Angular Momentum Constraints} \\
\hline
$0.5$ & $-2.42_{-5.52}^{+5.58}$& $-2.58_{-3.57}^{+3.48}$& $24.79_{-2.49}^{+2.32}$& $94.72_{-4.72}^{+4.85}$& $60.74_{-2.53}^{+2.62}$& $34.11_{-1.87}^{+1.77}$& $0.09_{-0.06}^{+0.06}$& $-0.06_{-0.06}^{+0.06}$& $-0.01_{-0.06}^{+0.06}$& $-0.59_{-0.01}^{+0.01}$& $0.16_{-0.01}^{+0.01}$\\
$1.5$ & $-1.38_{-6.76}^{+6.85}$& $-3.33_{-4.04}^{+3.88}$& $23.23_{-2.78}^{+3.03}$& $95.13_{-5.75}^{+5.39}$& $60.03_{-2.93}^{+2.92}$& $35.72_{-2.15}^{+1.99}$& $0.09_{-0.08}^{+0.07}$& $0.04_{-0.07}^{+0.07}$& $0.01_{-0.08}^{+0.07}$& $-0.59_{-0.01}^{+0.01}$& $0.15_{-0.01}^{+0.01}$\\
$2.5$ & $-4.97_{-7.75}^{+7.80}$& $-3.41_{-4.33}^{+4.20}$& $21.58_{-3.09}^{+3.16}$& $97.84_{-6.46}^{+6.60}$& $56.58_{-3.17}^{+3.31}$& $38.09_{-2.35}^{+2.33}$& $0.18_{-0.08}^{+0.08}$& $0.05_{-0.08}^{+0.08}$& $-0.01_{-0.08}^{+0.08}$& $-0.60_{-0.02}^{+0.02}$& $0.14_{-0.02}^{+0.02}$\\
$3.0$ & $-4.51_{-8.64}^{+8.42}$& $-4.21_{-4.85}^{+4.89}$& $20.82_{-3.50}^{+3.68}$& $95.67_{-7.29}^{+7.57}$& $55.17_{-3.61}^{+3.87}$& $39.33_{-2.60}^{+2.64}$& $0.19_{-0.09}^{+0.09}$& $0.03_{-0.09}^{+0.09}$& $-0.00_{-0.10}^{+0.09}$& $-0.61_{-0.02}^{+0.02}$& $0.14_{-0.02}^{+0.02}$\\
$3.5$ & $-5.62_{-9.79}^{+9.74}$& $-4.74_{-5.24}^{+5.75}$& $18.35_{-4.04}^{+4.42}$& $95.76_{-7.97}^{+8.52}$& $53.80_{-4.15}^{+4.41}$& $41.68_{-3.08}^{+3.17}$& $0.17_{-0.11}^{+0.11}$& $0.01_{-0.11}^{+0.10}$& $0.01_{-0.10}^{+0.11}$& $-0.62_{-0.02}^{+0.02}$& $0.14_{-0.02}^{+0.02}$\\
$4.0$ & $-1.29_{-11.86}^{+11.24}$& $-6.58_{-6.40}^{+6.26}$& $16.58_{-4.80}^{+5.22}$& $96.07_{-9.38}^{+9.32}$& $51.68_{-4.31}^{+4.82}$& $44.03_{-3.64}^{+3.83}$& $0.19_{-0.14}^{+0.13}$& $0.05_{-0.14}^{+0.12}$& $0.00_{-0.12}^{+0.11}$& $-0.62_{-0.03}^{+0.03}$& $0.15_{-0.03}^{+0.03}$\\
\hline
\multicolumn{12}{c}{Without Angular Momentum Constraints}\\
\hline
$2.5$ & $6.02_{-2.90}^{+2.78}$& $1.37_{-1.73}^{+1.69}$& $134.73_{-2.87}^{+2.74}$& $81.24_{-2.33}^{+2.39}$& $48.79_{-1.30}^{+1.34}$& $65.89_{-2.17}^{+2.12}$& $0.16_{-0.04}^{+0.04}$& $-0.09_{-0.04}^{+0.04}$& $0.02_{-0.04}^{+0.04}$& $-0.55_{-0.01}^{+0.01}$& $0.17_{-0.01}^{+0.01}$\\
$3.0$ & $4.10_{-4.03}^{+3.92}$& $0.79_{-2.33}^{+2.29}$& $114.24_{-3.61}^{+3.71}$& $85.80_{-3.16}^{+3.20}$& $50.45_{-1.82}^{+1.83}$& $71.63_{-2.65}^{+2.74}$& $0.21_{-0.05}^{+0.05}$& $-0.04_{-0.05}^{+0.05}$& $0.05_{-0.05}^{+0.05}$& $-0.57_{-0.01}^{+0.01}$& $0.16_{-0.01}^{+0.01}$\\
$3.5$ & $1.58_{-5.46}^{+5.70}$& $-1.17_{-3.04}^{+3.21}$& $95.67_{-5.20}^{+5.08}$& $90.80_{-4.42}^{+4.15}$& $52.65_{-2.48}^{+2.68}$& $77.05_{-3.74}^{+3.73}$& $0.27_{-0.06}^{+0.06}$& $-0.05_{-0.07}^{+0.07}$& $0.04_{-0.07}^{+0.06}$& $-0.59_{-0.02}^{+0.01}$& $0.16_{-0.01}^{+0.01}$\\
$4.0$ & $2.67_{-7.37}^{+7.61}$& $-3.38_{-4.36}^{+4.25}$& $74.88_{-6.60}^{+6.55}$& $93.99_{-6.05}^{+6.30}$& $54.09_{-3.42}^{+3.53}$& $78.41_{-5.05}^{+5.16}$& $0.28_{-0.08}^{+0.08}$& $-0.08_{-0.08}^{+0.09}$& $0.03_{-0.08}^{+0.08}$& $-0.63_{-0.03}^{+0.03}$& $0.18_{-0.03}^{+0.03}$\\
\enddata
}
\end{deluxetable*}

\begin{deluxetable*}{cccccccccccc}[h]
\label{table4}
\centering
\tablecaption{Same as Table~\ref{table3} but for the substructure component.}
\tablewidth{0pt}
\setlength{\tabcolsep}{1mm}{
\tablehead{
	\colhead{$z_{cut}$}&\colhead{$\mu_{r}$}&\colhead{$\mu_{\theta}$}&\colhead{$\mu_{\phi}$}&\colhead{$\sigma_{r}$}&\colhead{$\sigma_{\theta}$}&\colhead{$\sigma_{\phi}$}&\colhead{$\rho_{r\theta}$}&\colhead{$\rho_{r\phi}$}&\colhead{$\rho_{\theta\phi}$}&\colhead{$\mu_{\feh}$}&\colhead{$\sigma_{\feh}$}\\
\colhead{kpc}&\colhead{\kms}&\colhead{\kms}&\colhead{\kms}&\colhead{\kms}&\colhead{\kms}&\colhead{\kms}&\colhead{}&\colhead{}&\colhead{}&\colhead{dex}&\colhead{dex}}
\decimals
\startdata
\multicolumn{12}{c}{With Angular Momentum Constraints} \\
\hline
$0.5$ & $132.92_{-7.85}^{+7.78}$& $-5.34_{-3.92}^{+3.95}$& $3.93_{-2.23}^{+2.28}$& $110.25_{-5.10}^{+5.70}$& $78.50_{-2.93}^{+3.01}$& $42.41_{-1.49}^{+1.48}$& $0.10_{-0.06}^{+0.07}$& $0.05_{-0.06}^{+0.06}$& $-0.02_{-0.05}^{+0.05}$& $-1.22_{-0.04}^{+0.04}$& $0.38_{-0.03}^{+0.03}$\\
$1.5$ & $137.26_{-7.91}^{+8.21}$& $-4.59_{-4.26}^{+4.16}$& $3.89_{-2.34}^{+2.26}$& $109.69_{-5.20}^{+5.98}$& $76.73_{-3.22}^{+3.07}$& $42.96_{-1.56}^{+1.63}$& $0.13_{-0.07}^{+0.07}$& $-0.04_{-0.07}^{+0.06}$& $0.01_{-0.06}^{+0.06}$& $-1.23_{-0.04}^{+0.04}$& $0.38_{-0.04}^{+0.04}$\\
$2.5$ & $141.71_{-8.44}^{+8.54}$& $-6.08_{-4.64}^{+4.41}$& $3.62_{-2.76}^{+2.79}$& $107.35_{-5.54}^{+6.38}$& $76.50_{-3.54}^{+3.78}$& $44.59_{-1.87}^{+1.98}$& $0.18_{-0.07}^{+0.07}$& $-0.05_{-0.08}^{+0.08}$& $0.03_{-0.06}^{+0.06}$& $-1.25_{-0.04}^{+0.04}$& $0.37_{-0.04}^{+0.04}$\\
$3.0$ & $144.97_{-9.20}^{+9.13}$& $-8.35_{-5.14}^{+4.97}$& $3.64_{-3.11}^{+2.94}$& $105.25_{-5.59}^{+6.13}$& $77.34_{-3.91}^{+4.21}$& $45.75_{-2.21}^{+2.26}$& $0.20_{-0.08}^{+0.08}$& $-0.06_{-0.09}^{+0.09}$& $0.03_{-0.07}^{+0.07}$& $-1.28_{-0.06}^{+0.05}$& $0.39_{-0.05}^{+0.05}$\\
$3.5$ & $145.87_{-10.52}^{+10.52}$& $-9.78_{-5.61}^{+5.73}$& $3.13_{-3.56}^{+3.40}$& $105.07_{-6.10}^{+7.22}$& $78.23_{-5.43}^{+5.59}$& $46.93_{-2.57}^{+2.62}$& $0.20_{-0.08}^{+0.08}$& $-0.01_{-0.09}^{+0.09}$& $0.02_{-0.08}^{+0.08}$& $-1.32_{-0.07}^{+0.07}$& $0.42_{-0.06}^{+0.04}$\\
$4.0$ & $148.74_{-11.53}^{+14.90}$& $-10.77_{-6.90}^{+6.89}$& $4.02_{-4.52}^{+3.83}$& $106.73_{-7.79}^{+9.01}$& $79.52_{-7.01}^{+5.82}$& $48.81_{-3.05}^{+2.74}$& $0.21_{-0.10}^{+0.09}$& $0.00_{-0.10}^{+0.10}$& $0.03_{-0.09}^{+0.09}$& $-1.33_{-0.06}^{+0.07}$& $0.43_{-0.06}^{+0.04}$\\
\hline
\multicolumn{12}{c}{Without Angular Momentum Constraints}\\
\hline
$2.5$ & $120.09_{-6.61}^{+6.44}$& $-6.88_{-3.34}^{+3.35}$& $22.59_{-4.61}^{+4.74}$& $108.52_{-4.53}^{+5.20}$& $79.88_{-2.67}^{+2.64}$& $93.43_{-3.20}^{+3.20}$& $0.15_{-0.05}^{+0.05}$& $-0.08_{-0.06}^{+0.06}$& $0.06_{-0.04}^{+0.04}$& $-1.15_{-0.04}^{+0.04}$& $0.42_{-0.02}^{+0.02}$\\
$3.0$ & $131.37_{-7.21}^{+6.73}$& $-9.23_{-4.15}^{+3.84}$& $15.73_{-5.31}^{+5.32}$& $106.98_{-4.78}^{+5.31}$& $81.67_{-3.05}^{+3.07}$& $93.77_{-3.56}^{+3.56}$& $0.18_{-0.06}^{+0.06}$& $-0.09_{-0.06}^{+0.07}$& $0.08_{-0.05}^{+0.05}$& $-1.21_{-0.04}^{+0.04}$& $0.41_{-0.03}^{+0.02}$\\
$3.5$ & $134.31_{-7.87}^{+7.57}$& $-10.26_{-4.67}^{+4.93}$& $13.73_{-5.80}^{+5.52}$& $107.36_{-5.27}^{+6.02}$& $83.01_{-3.77}^{+3.89}$& $94.03_{-4.27}^{+3.95}$& $0.17_{-0.07}^{+0.07}$& $-0.10_{-0.07}^{+0.08}$& $0.08_{-0.05}^{+0.06}$& $-1.24_{-0.05}^{+0.04}$& $0.40_{-0.03}^{+0.03}$\\
$4.0$ & $144.29_{-9.84}^{+10.57}$& $-12.93_{-5.64}^{+5.53}$& $9.98_{-7.28}^{+7.19}$& $107.00_{-6.14}^{+7.01}$& $86.96_{-4.28}^{+4.58}$& $94.47_{-4.82}^{+4.89}$& $0.18_{-0.08}^{+0.07}$& $-0.11_{-0.09}^{+0.08}$& $0.10_{-0.07}^{+0.07}$& $-1.28_{-0.05}^{+0.05}$& $0.39_{-0.04}^{+0.04}$\\
\enddata
}
\end{deluxetable*}

\begin{deluxetable*}{cccccccccccc}[h]
\label{table5}
\centering
\tablecaption{Same as Table~\ref{table3} but for the halo component.}
\tablewidth{0pt}
\setlength{\tabcolsep}{1mm}{
\tablehead{
	\colhead{$z_{cut}$}&\colhead{$\mu_{r}$}&\colhead{$\mu_{\theta}$}&\colhead{$\mu_{\phi}$}&\colhead{$\sigma_{r}$}&\colhead{$\sigma_{\theta}$}&\colhead{$\sigma_{\phi}$}&\colhead{$\rho_{r\theta}$}&\colhead{$\rho_{r\phi}$}&\colhead{$\rho_{\theta\phi}$}&\colhead{$\mu_{\feh}$}&\colhead{$\sigma_{\feh}$}\\
\colhead{kpc}&\colhead{\kms}&\colhead{\kms}&\colhead{\kms}&\colhead{\kms}&\colhead{\kms}&\colhead{\kms}&\colhead{}&\colhead{}&\colhead{}&\colhead{dex}&\colhead{dex}}
\decimals
\startdata
\multicolumn{12}{c}{With Angular Momentum Constraints} \\
\hline
$0.5$ & $-10.30_{-18.62}^{+19.01}$& $-0.34_{-15.32}^{+15.60}$& $12.94_{-7.50}^{+7.15}$& $119.41_{-15.62}^{+15.22}$& $117.03_{-11.94}^{+15.41}$& $51.20_{-5.01}^{+6.43}$& $-0.04_{-0.14}^{+0.14}$& $-0.10_{-0.16}^{+0.15}$& $0.08_{-0.14}^{+0.13}$& $-2.03_{-0.06}^{+0.06}$& $0.21_{-0.04}^{+0.04}$\\
$1.5$ & $-11.23_{-21.51}^{+24.27}$& $0.36_{-17.95}^{+17.47}$& $14.23_{-8.38}^{+10.95}$& $119.92_{-20.57}^{+18.02}$& $115.16_{-11.84}^{+17.45}$& $53.88_{-6.08}^{+8.08}$& $-0.06_{-0.17}^{+0.15}$& $0.10_{-0.18}^{+0.19}$& $-0.04_{-0.15}^{+0.15}$& $-2.03_{-0.06}^{+0.08}$& $0.21_{-0.04}^{+0.06}$\\
$2.5$ & $-10.01_{-22.68}^{+21.43}$& $7.05_{-17.90}^{+18.42}$& $14.43_{-9.00}^{+10.15}$& $123.97_{-17.48}^{+17.27}$& $115.72_{-13.28}^{+18.78}$& $55.50_{-6.12}^{+8.21}$& $-0.05_{-0.16}^{+0.16}$& $0.11_{-0.17}^{+0.17}$& $-0.07_{-0.16}^{+0.16}$& $-2.03_{-0.07}^{+0.08}$& $0.22_{-0.05}^{+0.05}$\\
$3.0$ & $-11.83_{-32.01}^{+30.17}$& $6.98_{-24.40}^{+25.84}$& $16.08_{-11.39}^{+16.55}$& $125.16_{-22.83}^{+21.00}$& $121.02_{-17.87}^{+39.66}$& $59.75_{-8.21}^{+17.93}$& $-0.03_{-0.19}^{+0.20}$& $0.12_{-0.21}^{+0.23}$& $-0.04_{-0.18}^{+0.21}$& $-2.01_{-0.09}^{+0.31}$& $0.23_{-0.06}^{+0.22}$\\
$3.5$ & $-15.14_{-44.56}^{+43.37}$& $9.16_{-44.03}^{+44.69}$& $20.22_{-19.01}^{+31.10}$& $122.44_{-68.48}^{+34.90}$& $141.26_{-32.38}^{+51.70}$& $71.17_{-14.71}^{+36.97}$& $0.04_{-0.63}^{+0.29}$& $0.11_{-0.33}^{+0.42}$& $0.01_{-0.25}^{+0.32}$& $-1.96_{-0.15}^{+0.40}$& $0.29_{-0.12}^{+0.29}$\\
$4.0$ & $-17.50_{-47.63}^{+59.88}$& $14.71_{-54.04}^{+48.99}$& $22.82_{-24.25}^{+35.65}$& $126.39_{-56.79}^{+37.60}$& $153.86_{-41.19}^{+42.42}$& $77.91_{-17.39}^{+47.86}$& $0.08_{-0.55}^{+0.30}$& $0.02_{-0.36}^{+0.33}$& $0.06_{-0.28}^{+0.31}$& $-1.81_{-0.28}^{+0.28}$& $0.40_{-0.20}^{+0.21}$\\
\hline
\multicolumn{12}{c}{Without Angular Momentum Constraints}\\
\hline
$2.5$ & $-4.59_{-18.19}^{+16.53}$& $8.02_{-13.30}^{+13.74}$& $19.82_{-14.06}^{+14.19}$& $128.47_{-12.97}^{+12.85}$& $113.62_{-9.52}^{+11.59}$& $112.02_{-9.69}^{+11.00}$& $-0.07_{-0.12}^{+0.12}$& $0.03_{-0.13}^{+0.16}$& $-0.08_{-0.12}^{+0.12}$& $-2.04_{-0.06}^{+0.07}$& $0.21_{-0.04}^{+0.04}$\\
$3.0$ & $-7.41_{-20.85}^{+19.00}$& $16.01_{-14.27}^{+15.70}$& $22.18_{-16.36}^{+16.02}$& $127.30_{-14.47}^{+14.08}$& $111.46_{-10.46}^{+13.30}$& $113.48_{-10.97}^{+12.90}$& $-0.07_{-0.14}^{+0.14}$& $0.02_{-0.15}^{+0.16}$& $-0.10_{-0.13}^{+0.13}$& $-2.05_{-0.06}^{+0.07}$& $0.21_{-0.04}^{+0.04}$\\
$3.5$ & $-12.77_{-25.26}^{+24.33}$& $18.86_{-19.23}^{+20.68}$& $19.36_{-18.59}^{+21.11}$& $132.05_{-17.40}^{+16.77}$& $112.32_{-12.67}^{+18.44}$& $117.99_{-13.23}^{+18.75}$& $-0.01_{-0.17}^{+0.18}$& $-0.03_{-0.17}^{+0.19}$& $-0.12_{-0.15}^{+0.15}$& $-2.08_{-0.07}^{+0.08}$& $0.20_{-0.04}^{+0.06}$\\
$4.0$ & $-15.40_{-30.43}^{+28.04}$& $19.34_{-21.91}^{+25.40}$& $22.07_{-22.28}^{+24.34}$& $130.62_{-20.40}^{+20.83}$& $112.94_{-16.24}^{+19.77}$& $115.40_{-15.42}^{+18.53}$& $0.06_{-0.21}^{+0.23}$& $0.03_{-0.21}^{+0.23}$& $-0.15_{-0.18}^{+0.17}$& $-2.10_{-0.07}^{+0.08}$& $0.19_{-0.04}^{+0.05}$\\
\enddata
}
\end{deluxetable*}

\bibliography{paper_draft}
\bibliographystyle{aasjournal}
\end{document}